\documentclass[useAMS,usenatbib]{mn2e}
\usepackage{times}
\usepackage{epsfig}
\usepackage{graphicx}
\usepackage{color}

\def\aj{AJ}
\def\pasp{PASP}
\def\apj{ApJ}
\def\apjs{ApJS}
\def\apjl{ApJL}
\def\aap{A\&A}
\def\araa{ARA\&A}

\def\mnras{MNRAS}

\def\apss{Astrophysics and Space Science}

\newcommand{\iso}[2]{\ensuremath{^{#1}\rm{#2}}}

\def\mesa{{\sc mesa}}
\def\kepler{{\sc kepler}}
\def\cmfgen{{\sc cmfgen}}
\def\v1d{{\sc v1d}}

\def\msun{M$_{\odot}$}
\def\lsun{L$_{\odot}$}
\def\rsun{R$_{\odot}$}

\def\one{{\,\sc i}}
\def\two{{\,\sc ii}}
\def\three{{\,\sc iii}}
\def\four{{\,\sc iv}}
\def\five{{\sc v}}
\def\six{{\sc vi}}

\def\cm3{cm$^{-3}$}
\def\kms{km~s$^{-1}$}

\def\ergs{erg\,s$^{-1}$}

\def\lesssim{\mathrel{\hbox{\rlap{\hbox{\lower4pt\hbox{$\sim$}}}\hbox{$<$}}}}
\def\gtrsim{\mathrel{\hbox{\rlap{\hbox{\lower4pt\hbox{$\sim$}}}\hbox{$>$}}}}

\title[A study of SN\,2008bk]{A study of the low-luminosity Type II-Plateau supernova 2008bk}
\author[S.~M.~Lisakov et al.]
{S.~M.~Lisakov,$^{1}$\thanks{E-mail: lisakov57@gmail.com}
  Luc Dessart,$^1$
D. John Hillier,$^{2}$
Roni Waldman,$^3$
and Eli Livne$^3$
  \\
  \\
  $^{1}$: Laboratoire Lagrange, UMR7293, Universit\'e Nice Sophia-Antipolis, CNRS,
  Observatoire de la C\^{o}te d'Azur, 06304 Nice, France. \\
  $^2$: Department of Physics and Astronomy \& Pittsburgh Particle Physics,
Astrophysics, and Cosmology Center (PITT PACC),  University of Pittsburgh, \\
3941 O'Hara Street, Pittsburgh, PA 15260, USA. \\
$^3$: Racah Institute of Physics, The Hebrew University, Jerusalem 91904, Israel. \\
}

\begin{document}

\date{Accepted . Received }

\pagerange{\pageref{firstpage}--\pageref{lastpage}} \pubyear{2016}
\maketitle
\label{firstpage}

\begin{abstract}
Supernova (SN) 2008bk is a well observed low-luminosity Type II event visually
associated with a low-mass red-supergiant progenitor.
To model SN\,2008bk, we evolve a 12\,\msun\ star from the main sequence until core collapse, when
it has a total mass of 9.88\,\msun, a He-core mass of 3.22\,\msun, and a radius of 502\,\rsun.
We then artificially trigger an explosion that produces 8.29\,\msun\ of ejecta with a total energy
of 2.5$\times$10$^{50}$\,erg  and $\sim$\,0.009\,\msun\ of \iso{56}Ni.
We model the subsequent evolution of the ejecta with
non-Local-Thermodynamic-Equilibrium time-dependent radiative transfer.
Although somewhat too luminous and energetic, this model reproduces satisfactorily the
multi-band light curves and multi-epoch spectra of SN\,2008bk, confirming the suitability of
a low-mass massive star progenitor.
As in other low-luminosity SNe II, the structured H$\alpha$ profile at the end of the plateau phase
is probably caused by Ba\two\,6496.9\,\AA\ rather than asphericity.
We discuss the sensitivity of our results to changes in progenitor radius and mass, as well as chemical mixing.
A 15\% increase in progenitor radius causes a 15\% increase in luminosity and a 0.2\,mag $V$-band brightening
of the plateau but leaves its length unaffected. An increase in ejecta mass by 10\% lengthens the plateau by $\sim$\,10\,d.
Chemical mixing introduces slight changes to the bolometric light curve, limited to the end of the plateau,
but has a large impact on  colours and spectra at nebular times.
\end{abstract}


\section{Introduction}

Massive stars with an initial mass greater than $\sim$\,8~\msun\ are expected to end their lives
with the gravitational collapse of their degenerate core.
If a successful explosion follows, an H-rich progenitor leads to a Type II supernova (SN),
the most frequently observed type of core-collapse SNe \citep{smith_11_sn_stat}.
These SNe are characterised by the presence of strong hydrogen Balmer lines in their optical spectra.
Their high-brightness phase, which also coincides with the photospheric
phase, lasts about one hundred days. Early simulations of such Type II SNe 
(\citealt{grassberg_71}, \citealt{FA77})
suggest a progenitor star with a massive and extended H-rich envelope, as typically found in red-supergiant (RSG) stars.
The physics underlying the evolution of a SN II-Plateau (II-P) has been extensively discussed by, 
for example, \cite{FA77} or more recently \citet{utrobin_99em_07}.
The association between Type II-P SNe and RSG stars has also been more directly made through
the identification of the progenitor star on pre-explosion images (see \citealt{smartt09} for a review).

Type II SNe exhibit a broad range in $R$-band absolute magnitude, spanning about 5 mag during
the photospheric phase
and the nebular phase \citep{hamuy_03}. The latter suggests a range of $\gtrsim$\,10 in the mass
of  \iso{56}Ni ejected. The ejecta expansion rate inferred from P-Cygni profile widths halfway
through the plateau
also spans a range from 1000 to 8000\,\kms\  \citep{hamuy_03}, suggestive of a large
scatter in the ratio of ejecta kinetic energy $E_{\rm kin}$ and ejecta mass $M_{\rm e}$.
More recent surveys document this diversity further
\citep{anderson_2pl,faran_sn2p_14,faran_sn2l_14,sanders_sn2_15,galbany_sn2_15}, and also emphasize,
for example, the range in photospheric phase durations, the $V$-band decline rate after maximum,
how various radiative properties correlate.

The existence of low-luminosity (i.e., intrinsically faint) SNe II was well established in the 90s.
SN\,1997D was one of the first Type II to exhibit abnormally narrow P-Cygni profiles (of the order of 1000\,\kms)
and a low optical brightness during the photospheric phase (always fainter than $-$14.65\,mag in the $V$-band;
\citealt{turatto_97d_98}), both suggestive of a low energy explosion in an extended H-rich star.
The sample of low-luminosity Type II SNe now includes a handful of objects \citep{pastorello04,spiro14}, with
an absolute  $V$-band magnitude that covers from $-$14 to $-$15.5\,mag around 50\,d after explosion,
thus 2$-$\,3\,mag fainter
than the prototypical Type II-P  SN\,1999em \citep{leonard_99em_02}.
The fraction of low-luminosity SNe could be $\sim$\,5\% of all Type II SNe \citep{pastorello04}.

Numerous radiation-hydrodynamic simulations of low-luminosity Type II SNe have been carried out.
Constraints on the progenitor star and the ejecta (explosion energy, \iso{56}Ni mass) are obtained
through iteration until a good match to the bolometric light curve and the
photospheric velocity evolution.
\citet{utrobin_03z_07} modelled the low-luminosity SN\,2003Z and obtained an ejecta mass
of 14\,\msun, an ejecta kinetic energy of 2.45$\times$10$^{50}$\,erg, 0.006\,\msun\ of \iso{56}Ni,
and a progenitor radius of 230\,\rsun. \citet{spiro14} used a similar approach and proposed
progenitor masses in the range 10--15\,\msun\ for the whole sample of low-luminosity SNe II.
These results suggest that low-luminosity Type II SNe are intrinsically under-energetic and synthesize little \iso{56}Ni.
Other studies have argued that these low-luminosity Type II SNe are the result of weak explosions
in higher mass RSG stars. In this context, the low \iso{56}Ni mass arises from the significant
fallback of material onto the proto-neutron star, associated potentially with the subsequent formation
of a black hole. 

 \citet{turatto_97d_98} propose this scenario with a 26\,\msun\ progenitor star
for SN\,1997D \citep{benetti_97D_01}, and argue against the possibility of a lower mass progenitor.
Later work by \citet{zampieri_03}, using a semi-analytic modelling of the light curve,
give support to the association with higher mass progenitors, in which the explosion
is followed by fallback.
The situation is therefore  unsettled. For SN\,1997D, \citet{zampieri_07} proposes a 14\,\msun\ progenitor,
which disagrees with \citet{turatto_97d_98}. For the low-luminosity SNe II-P 2005cs and 2008bk,
a progenitor detection exists and suggests a low/moderate mass massive star
(in the range 9--13\,\msun\ on the zero-age main-sequence, ZAMS; \citealt{maund_05cs_05};
 \citealt{li_05cs_06}; \citealt{mattila_08bk_08}; \citealt{vandyk_08bk_12}; \citealt{maund_08bk_14}
 ).

Even when an ejecta mass is inferred from light-curve modelling, estimating the corresponding progenitor mass
on the main sequence is subject to error because of the uncertain
mass loss history. Furthermore, the light curve modelling above is primarily sensitive to
the H-rich ejecta mass, not the total ejecta mass. The helium core mass can only be estimated
by modelling nebular phase spectra and the helium core dynamics
\citep{DLW10b,DH11,maguire_2p_12,jerkstrand_etal_12,dessart_etal_13}.

In this work, we model the low-luminosity SN\,2008bk because it is the best observed SN of this
class of event, with a good photometric monitoring in the optical that started about a week
after explosion \citep{pignata_08bk_13}.
Based on pre-explosion images and evidence of the disappearance of
a source on post-explosion images, inferences have been made to constrain
the nature of the progenitor star and its mass. The consensus is that it is a RSG
star, although its inferred ZAMS mass differs somewhat between studies,
with 8--8.5\,\msun\ (\citealt{mattila_08bk_08,vandyk_08bk_12}; see also \citealt{vandyk_08bk_13})
and 11.1--14.5\,\msun\ \citep{maund_08bk_14}.
In contrast,  there has been little analysis of the SN spectroscopic and photometric data.
A preliminary analysis is presented in \citet{pignata_08bk_13}, who proposes an ejecta with
$E_{\rm kin}=$\,2.5$\times$10$^{50}$\,erg, a total mass of 12\,\msun, and a \iso{56}Ni mass of 0.009\,\msun,
together with a progenitor radius of 550\,\rsun,
\citet{maguire_2p_12} present an analysis of a nebular-phase spectrum at 547\,d after explosion
and propose a progenitor star with a main-sequence mass of 12\,\msun.

The simulations presented here are based on models of the progenitor evolution from the main-sequence
until core collapse, together with the subsequent simulation of the piston-driven explosion including
explosive nucleosynthesis. The bulk of the work lies, however, in the non-Local-Thermodynamic-Equilibrium
(nLTE) time-dependent
radiative-transfer modelling of the photometric and spectroscopic evolution of SN\,2008bk.
Since a low/moderate mass progenitor has been proposed by all former studies on SN\,2008bk,
we limit our investigation to a progenitor star of 12\,\msun\ on the main sequence.
In a forthcoming study, we will investigate the properties of the whole sample of low-luminosity Type II SNe,
and consider progenitors from both low and high mass RSG stars.

In the next section, we summarise the source of observational data for SN\,2008bk.
In Section~\ref{sect_model}, we present our numerical approach for the modelling of the
pre-SN evolution with \mesa\ \citep{mesa1,mesa2,mesa3},
the explosion with \v1d\ \citep{livne_93,DLW10a, DLW10b},
and the modelling of the SN radiation from 10\,d after explosion until nebular times
with \cmfgen\ \citep{HM98,DH05a,DH08, HD12,dessart_etal_13}.
Our best match model to the photometric and spectroscopic observations of SN\,2008bk is
presented in Section~\ref{sect_X}.
In Section~\ref{sect_dep}, we discuss the sensitivity of our results to changes
in progenitor/ejecta radius (Section~\ref{sect_rad}), progenitor mass (Section~\ref{sect_mass}),
and chemical mixing (Section~\ref{sect_mixing}).
We conclude in Section~\ref{sect_conclusion}.


\section{Observational data}
\label{sect_data}

The optical and near-IR photometry that we use for SN\,2008bk comes from  \citet{pignata_08bk_13}.
For the spectroscopic data, we limit our analysis to the optical range, and use observations
from Pignata as well as the spectropolarimetric observations of \citet[see also Leonard et al., in prep.]{leonard_08bk_12}.
The spectropolarimetric data is, however, not accurately calibrated in flux (no flux standard was used during
the observing night) so the corresponding spectra are used mostly to compare the morphology of
line profiles (in practice, we distort the observed spectrum so that it has the same overall shape as
the model at the same epoch; see Section~\ref{sect_X}).

Following the earlier estimate of a low reddening towards SN\,2008bk \citep{pignata_08bk_13},
we adopt an $E(B-V)=$\,0.02\,mag. This is within 0.01\,mag of the value reported
by \citet{schlegel_ebv_98} for the line-of-sight towards NGC\,7793, the galaxy host of SN\,2008bk.
This suggests the source of extinction to SN\,2008bk is exclusively galactic.
Our reference model X (see below) gives a satisfactory agreement to the spectral and colour
evolution of SN\,2008bk
but it overestimates its plateau brightness. This discrepancy is reduced with a larger reddening
although an $E(B-V)$ value greater than $\sim$\,0.1\,mag causes a mismatch in colour.

We adopt the Cepheid-based distance modulus of 27.68$\pm$\,0.05\,mag (internal error) $\pm$\,0.08\,mag
(systematic error) inferred by \citet{sn08bk_cepheid_10}.
Finally, we adopt a recession velocity of 283\,\kms\ for SN\,2008bk, as inferred from nearby H\two\ regions
(Pignata, private comm.).

By drawing an analogy between the light curves of SN\,2008bk and SN\,2005cs,
\citet{pignata_08bk_13} proposes an explosion date of MJD\,54548.0$\pm$\,2.
Our best match model yields an improved agreement with observations if we adopt
instead MJD\,54546.0, which is within the uncertainty of Pignata's choice.


\begin{figure}
  \includegraphics[width=0.49\textwidth]{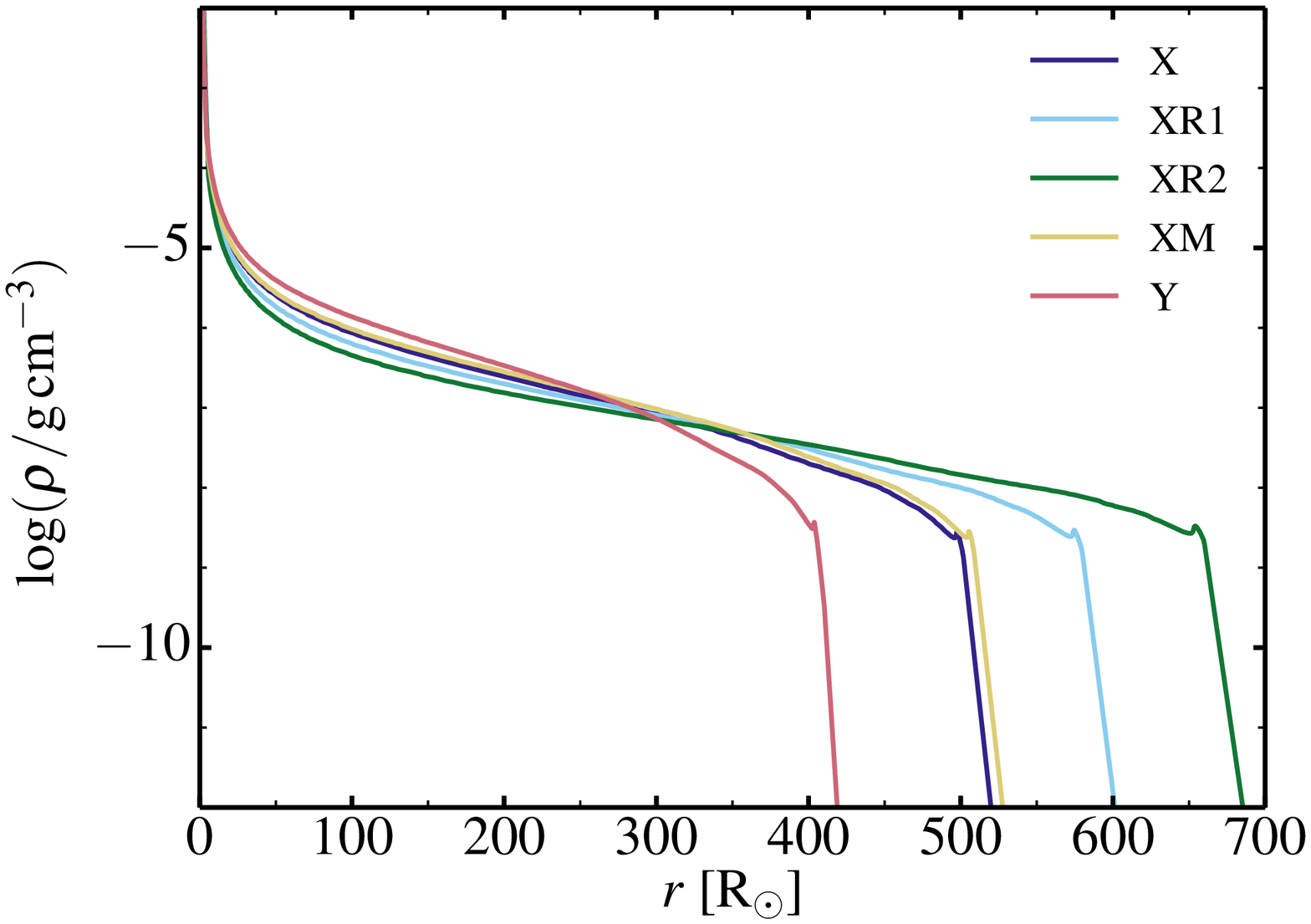}
  \includegraphics[width=0.49\textwidth]{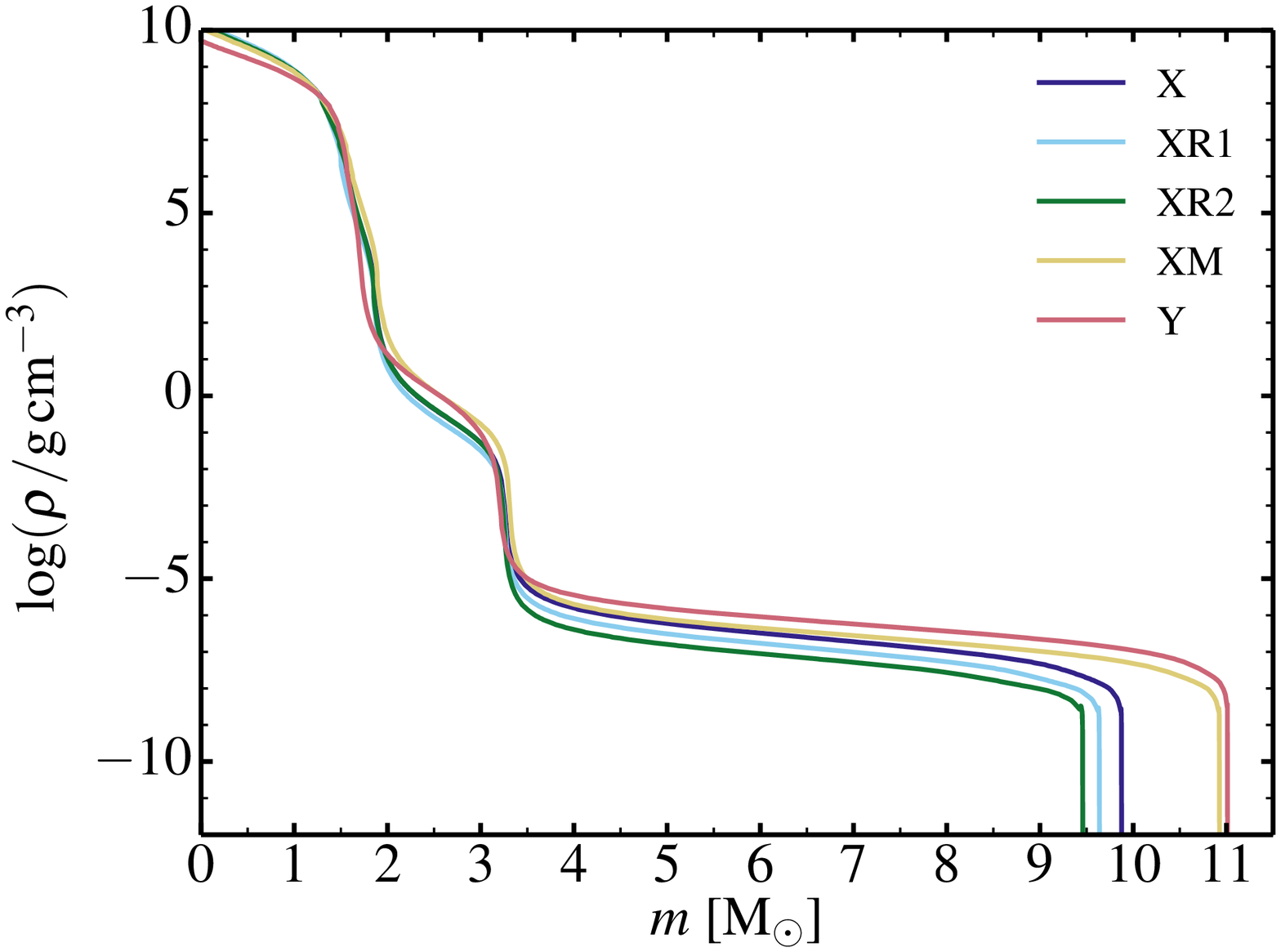}
  \includegraphics[width=0.49\textwidth]{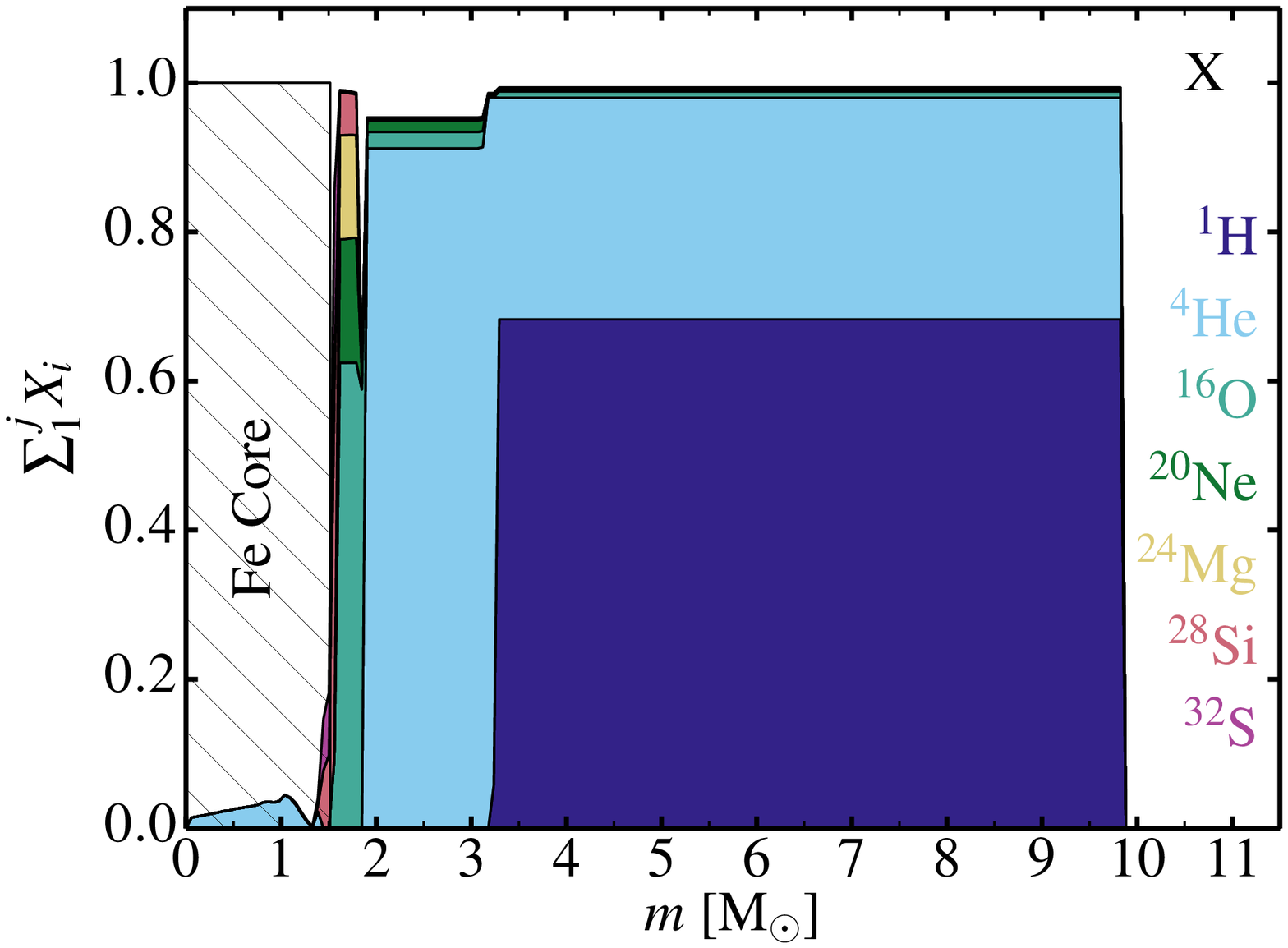}
\caption{Density structure versus radius (top) and mass (middle) for our set of pre-SN models,
and composition profile versus mass for model X (bottom). For the latter, we show the cumulative
mass fraction to better highlight the dominance of H and He in such a low-mass massive-star model.
The Si-rich and the O-rich shells occupy a very narrow mass range.
\label{fig_prog}
}
\end{figure}

\begin{figure}
  \includegraphics[width=0.48\textwidth]{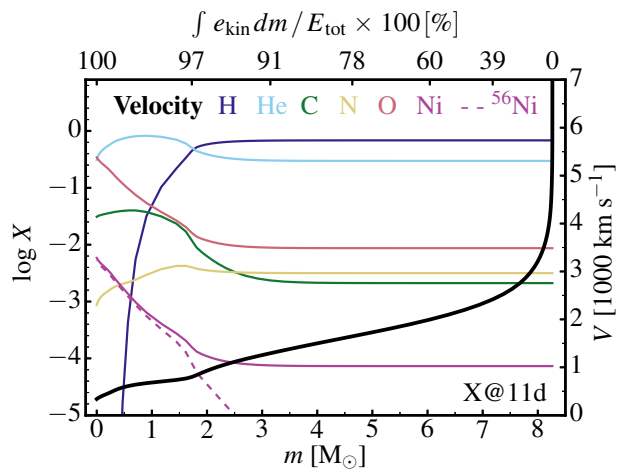}
  \caption{Variation of the mass fraction for H, He, C, N, O, and Ni (the dashed
  line gives the contribution from the \iso{56}Ni) with interior mass and
  velocity (thick line; see right axis) for model X at 11\,d.
  We have applied some chemical mixing for all species (see Section~\ref{sect_v1d} for details).
   The top axis shows the depth variation of the fractional inward-integrated kinetic energy.
   About 50\% of the total ejecta kinetic energy is contained in the outer 1\,\msun\ of the ejecta,
   and only a few per cent in the former He core (below 1000\,\kms).
  \label{fig_X_comp}}
\end{figure}

\section{Numerical setup}
\label{sect_model}

\subsection{Pre-supernova evolution}
\label{sect_mesa}

   Using in all cases an initial mass of 12\,\msun, we evolved several models
   from the main sequence to the gravitational collapse
   of the iron core using \mesa\ version 7623 \citep{mesa1,mesa2,mesa3}.
   We used the default massive star parameters of \mesa\ (those given in
   {\tt inlist\_massive\_defaults}), except for the adjustments discussed below.

   Models X, XR1, XR2, and XM were evolved at a metallicity of 0.02.
   Convection is followed according to the Ledoux criterion,
   with a mixing length parameter $\alpha_{\rm MLT}$ that varies between models,
   a semi-convection efficiency parameter $\alpha_{\rm sc}=$\,0.1, and an exponential overshoot
   with parameter $f=$\,0.004.
   The mixing length parameter
   $\alpha_{\rm MLT}$ was adjusted to modulate the stellar radius at the time of death.
   In \citet{dessart_etal_13}, we found that to reconcile the colour evolution of SNe II-P
   with observations, more compact RSG progenitors were needed. In \mesa, this can be achieved
   by employing for $\alpha_{\rm MLT}$ a value larger than standard (the `standard' used by default in \mesa\ is 2.0) 
   --- by enhancing the efficiency of convection within the MLT
   formalism we allow the star to carry the core radiative flux through a denser/smaller envelope.
   Models X, XR1, and XR2 use a value of $\alpha_{\rm MLT}$
   of 3, 2.5, and 2, respectively. For these three models, we enhanced the mass loss rate from the
   ``Dutch" recipe by a factor of 2.5. This choice is to compensate for the lower mass loss rates that result
   from the higher effective temperatures of our more
   compact RSG models and also to reduce the final star mass to $\lesssim$\,10\,\msun\ (originally
   motivated by the low progenitor masses of 8-8.5\,\msun\ inferred by \citealt{mattila_08bk_08}
   and \citealt{vandyk_08bk_12}).
   In model XM, we use $\alpha_{\rm MLT}=$\,3, but enhance the mass loss by a factor of 1.5 instead of 2.5 in order
   to obtain a higher mass model at death.

   Model X is our best match model to the observations of SN\,2008bk so these other models
   are computed to explore the sensitivity of our results to a number of evolution parameters.
   In a 12\,\msun\ star, most of the mass loss occurs during the RSG phase. Hence, the modulations in
   the mass loss rate impact primarily the H-rich envelope mass and leave the He-core largely unaffected.
   As discussed later, 10\% changes in H-rich envelope mass or progenitor radius can visibly impact
   the resulting SN radiation, although in different ways.

   Model Y was computed as part of a separate effort (with \mesa\ version 4723)
   and so uses slightly different parameters from the above set.
   It was evolved at  a metallicity of 0.0162 with the `Dutch' mass-loss recipe,
   with an enhancement factor of 1.4.
   Convection is followed according to the Ledoux criterion,
   with a mixing length parameter $\alpha_{\rm MLT}=$\,3,
   a semi-convection efficiency parameter $\alpha_{\rm sc}=$\,0.1, and an exponential overshoot
   with parameter $f=$\,0.008.
   Model Y is included here to discuss the effect of mixing (see next section) on the SN ejecta and radiation.

   These various incarnations of a 12\,\msun\ star yield similar stellar properties at the time
   of core collapse. All models produce RSG stars at death with surface radii of 405--661\,\rsun,
   total masses of 9.46--11.01\,\msun, and He-core masses of 3.13--3.26\,\msun.
   The progenitor density structure at collapse shows the typical RSG profile, with a high-density highly bound He core
   and a low-density extended massive H-rich envelope (Fig.~\ref{fig_prog}).
   In such a low-mass massive star, there is only $\sim$\,0.2\,\msun\ between the outer edge of
   the iron core and the inner edge of the He-rich shell. The H-rich shell is about 7\,\msun, hence about 7 times
   more massive than the He-rich shell. \\

\begin{table*}
\caption{
Summary of the progenitor properties at the onset of iron-core collapse and of the ejecta properties.
All models start on the main sequence with a mass of 12\,\msun\ and a metallicity $Z$.
Model X is our best-match model to the observations of SN\,2008bk.
Other models are used to discuss the sensitivity of our results to variations
in progenitor radius (XR1 and XR2), progenitor mass (XM),
or \iso{56}Ni mixing (YN1, YN2, and YN3).
For the iron core, we report the location where the electron fraction rises outward to 0.499,
for the CO core where the oxygen mass fraction rises inward to 0.1, and the He core where the hydrogen
mass fraction drops inward to 0.001 --- the exact criterion for these masses is unimportant.
The column \iso{56}Ni$_0$ gives the original total mass of \iso{56}Ni.
The mass of the H-rich envelope is $M_\star - M_{\rm He-core}$ and covers the range $\sim$\,6.3--7.9\,\msun\ in
the progenitor star. When applying mixing, H may be mixed into shells that were originally H deficient
so  the total mass of H-rich shells (i.e., with $X_{\rm H}>$\,0.1) may shift to a larger value in the corresponding 
ejecta --- this does not apply to models YN1, YN2, and YN3 in which we mix only \iso{56}Ni.
Numbers in parentheses correpond to powers of ten.
See Section~\ref{sect_model} for discussion.
\label{tab_models}
}
\begin{tabular}{l@{\hspace{3mm}}c@{\hspace{3mm}}
c@{\hspace{3mm}}c@{\hspace{3mm}}
c@{\hspace{3mm}}c@{\hspace{3mm}}c@{\hspace{3mm}}
c@{\hspace{3mm}}c@{\hspace{10mm}}c@{\hspace{3mm}}
c@{\hspace{3mm}}c@{\hspace{3mm}}c@{\hspace{3mm}}
c@{\hspace{3mm}}c@{\hspace{3mm}}c@{\hspace{3mm}}}
\hline
Model &
Z &
$R_\star$  &
$M_\star$   &
$T_\star$  &
$L_\star$   &
Fe core &
CO core &
He core &
Ejecta &
H-rich   &
H   &
He   &
O   &
\iso{56}Ni$_0$  &
$E_{\rm kin}$   \\
\hline
  & & [\rsun] & [\msun] & [K] & [\lsun] & [\msun]  & [\msun]& [\msun]  & [\msun]  &[\msun]  &[\msun]  &[\msun]  &[\msun]  &[\msun]  & [10$^{50}$\,erg] \\
  \hline
X     & 0.02 &502  & 9.88  &3906  &52733  & 1.53  &1.86 & 3.22 &8.29  & 6.66   &4.54  &3.24  &0.22  & 8.57(-3)  & 2.5      \\ 
XR1   & 0.02   &581  & 9.63  &3644  &53436   & 1.50 & 1.85 & 3.20 & 8.08  & 6.43   & 4.38  &3.17  &0.19  & 8.19(-3)  &2.6     \\ 
XR2     & 0.02 &661  & 9.45  &3390  & 51770   & 1.60 & 1.85 & 3.19  & 7.90  &6.26   &4.25  &3.13  &0.22  & 9.00(-3)  &2.7     \\ 
\hline
X        & 0.02 &502  & 9.88  &3906  &52733  & 1.53  &1.86 & 3.20  &8.29  & 7.10   &4.54  &3.24  &0.22  &  8.57(-3) & 2.5      \\  
XM     & 0.02 &510  &10.92  &3943  &56408  & 1.50 &1.89 & 3.27  & 9.26  & 7.65   & 5.20  &3.54  &0.23  & 7.20(-3)  &2.7     \\ 
\hline
YN1       & 0.0162   &405  &11.01  &4195  &45715  & 1.38 & 1.73 & 3.15 &9.45  &7.86   &5.41  &3.74  &0.09  & 1.00(-2)  &2.5    \\ 
YN2       & 0.0162   &405  &11.01  &4195  &45715  & 1.38 & 1.73 & 3.15 &9.45  &7.86   &5.41  &3.74  &0.09  & 1.00(-2)  &2.5    \\ 
YN3       & 0.0162   &405  &11.01  &4195  &45715  & 1.38 & 1.73 & 3.15 &9.45  &7.86   &5.41  &3.74  &0.09  &  1.00(-2) &2.5    \\ 
\hline
\end{tabular}
\end{table*}

\subsection{Piston-driven explosions}
\label{sect_v1d}

   The \mesa\ simulations were stopped when the maximum infall core velocity reached 1000\,\kms.
   At that time, we remapped the \mesa\ model into \v1d\ \citep{livne_93,DLW10a,DLW10b}.
   The model was resampled onto
   a grid with a mass resolution $\delta m$ of $10^{-4}-10^{-3}$\,\msun\ at the base,
   increasing to 10$^{-2}$\,\msun\ at and beyond 2\,\msun. Within a few percent of the stellar surface,
   the mass resolution is progressively increased to have a surface resolution of $10^{-6}-10^{-5}$\,\msun.
   At the progenitor surface, we go down to a density of $10^{-12}$\,g\,cm$^{-3}$ but ideally one should
   use an even lower value in order to have optically thin shells at the outer boundary where the shock breaks out.

    The explosion is in all cases triggered by moving a piston at $\sim$\,10,000\,\kms\ at the inner boundary,
    which we place at the location where the entropy rises outward from the center to 4\,k$_{\rm B}$\,baryon$^{-1}$
    (see, e.g., \citealt{ugliano_ccsn_12}). This location is typically in the outer
    part of the Si-rich shell, just below the O-rich shell, and located around 1.55\,\msun\ in these models.
    Because the explosion energy is low in our SN\,2008bk models, some fallback of the order of 0.01\,\msun\ 
    may occur.

    The explosion models were done iteratively until we obtained an ejecta in homologous expansion with
    the desired \iso{56}Ni mass of $\sim$\,0.009\,\msun\ and total energy of 2.5$\times$10$^{50}$\,erg
    \citep{pignata_08bk_13}. Iteration is needed because the \iso{56}Ni mass is sensitive
    to the piston properties (location, speed) and to the magnitude of fallback. The asymptotic ejecta
    energy also depends on fallback. 

      In most of our simulations, we enforce a chemical mixing using a boxcar algorithm (see \citealt{dessart_etal_12}
     for discussion) that affects all species. We also explore the impact
     of mixing only \iso{56}Ni (and substituting it with H to keep the mass fraction normalised to unity at each depth).
     Three levels of mixing are enforced, increasing from model YN1 to YN2 and YN3. These simulations
     and additional information on the mixed composition profiles are presented in Section~\ref{sect_mixing}.

     The ejecta properties of our model set are given in Table~\ref{tab_models}.
     We also show the composition profile for model X in mass and velocity space
     in Fig.~\ref{fig_X_comp}.

     \begin{figure}
  \includegraphics[width=0.48\textwidth]{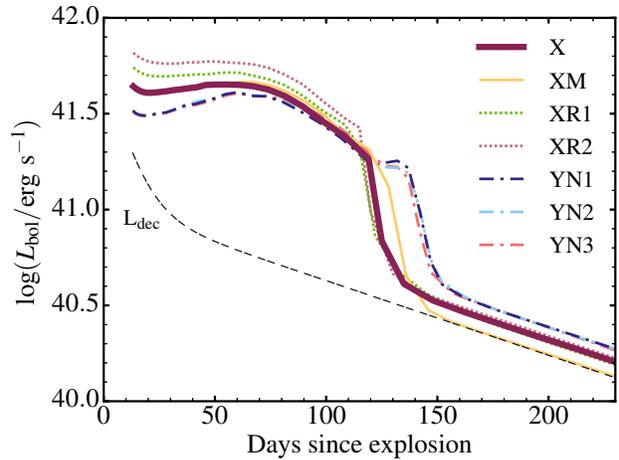}
    \caption{
    Bolometric light curves computed with \cmfgen\ for our set of models.
    Model X is our best-match model to the observations of SN\,2008bk.
    Our sample includes models that differ only slightly from model X,
    either in the pre-SN evolution or in the explosion properties, in order to test the sensitivity of our results to changes
    in $R_\star$ (models X, XR1, and XR2), changes in H-rich envelope mass (model X and XM), and changes
    in \iso{56}Ni mixing (models YN1, YN2, YN3) --- see
    Table~\ref{tab_models} for details.
    The dashed line corresponds to the instantaneous decay power from an initial mass of 0.0072\,\msun\ of \iso{56}Ni.
    \label{lc_lbol}
    }
\end{figure}

\begin{table*}
\caption{
Sample of results for our set of simulations. Model X is the closest match to the observations
of SN\,2008bk, while other models are used to test the sensitivity of our results to changes in
progenitor and explosion characteristics (see also Table~\ref{tab_models}).
Here, we give the approximate duration $\Delta t_{\rm P}$ of the ``plateau" phase (we set its end
when the bolometric luminosity or $V$-band magnitude suddenly drops),
and then the bolometric luminosity, the absolute $V$-band magnitude, the $(U-V)$ colour,
the photospheric velocity (the electron-scattering opacity is used),
and the velocity at maximum absorption in H$\alpha$ at 15
and 50\,d after explosion. Numbers in parentheses correpond to powers of ten.
\label{tab:res}
}
\begin{tabular}{l@{\hspace{3mm}}c@{\hspace{3mm}}
c@{\hspace{3mm}}c@{\hspace{3mm}}c@{\hspace{3mm}}c@{\hspace{3mm}}
c@{\hspace{3mm}}c@{\hspace{3mm}}c@{\hspace{3mm}}c@{\hspace{3mm}}
c@{\hspace{3mm}}c@{\hspace{3mm}}c@{\hspace{3mm}}c@{\hspace{3mm}}}
\hline
Model  &
$E_{\rm kin}$       &
\iso{56}Ni$_0$  &
$\Delta t_{\rm P}$  &
\multicolumn{2}{c}{$L_{\rm bol}$ }  &
\multicolumn{2}{c}{$M_V$}   &
\multicolumn{2}{c}{$U-V$ }    &
\multicolumn{2}{c}{$V_{\rm phot}$ } &
\multicolumn{2}{c}{$V_{\rm abs}$(H$\alpha$) } \\
\hline
 &
 [10$^{50}$\,erg]   &
[\msun]  &
[d] &
 [\ergs] & [\ergs] &
[mag] & [mag] &
[mag] & [mag] &
[\kms] & [\kms] &
[\kms] & [\kms] \\
\hline
&
&
&
&
(15\,d)& (50\,d)&
(15\,d)& (50\,d)&
(15\,d)& (50\,d)&
(15\,d)& (50\,d)&
(15\,d)& (50\,d) \\
\hline
\hline
X      &2.5       &  8.57(-3)   &120         &4.36(41)  &4.46(41)  &--15.46    &--15.50    &--0.01     &2.11      &4833  &2334   & 5362  & 4208 \\
XR1    &2.6       &  8.19(-3)  &115       &5.32(41)  &5.13(41)  &--15.60    &--15.67    &--0.33     &1.96      &4854   &2468   &5344      &4251 \\
XR2    &2.7       &  9.00(-3)  &115       &6.33(41)  &5.92(41)  &--15.70    &--15.84    &--0.56     &1.79      &4685  &2612   &4887      &3979 \\
\hline
X      &2.5       & 8.57(-3)  &120       &4.36(41)  &4.46(41)  &--15.46    &--15.50    &--0.01     &2.11      &4833   &2334   & 5362  & 4208  \\
XM     &2.7       &  7.20(-3)  &120       &4.23(41)  &4.53(41)  &--15.43    &--15.51    &--0.01     &2.10      &4731   &2323   &5126     &4101 \\
\hline
YN1    &2.5       & 1.00(-2)   &140       &3.16(41)  &3.81(41)  &--15.16    &--15.30    &0.41      &2.29      &4534   &2177   &5057      &4197 \\
YN2    &2.5       & 1.00(-2)   &140       &3.15(41)  &3.81(41)  &--15.16    &--15.30    &0.41      &2.31      &4530   &2182   &5072      &4200 \\
YN3    &2.5       &  1.00(-2)    &140       &3.13(41)  &3.75(41)  &--15.16    &--15.27    &0.44      &2.49      &4519   &2194   &5042      &4555 \\
\hline
\end{tabular}
\end{table*}

\begin{figure*}
  \includegraphics[width=0.48\textwidth]{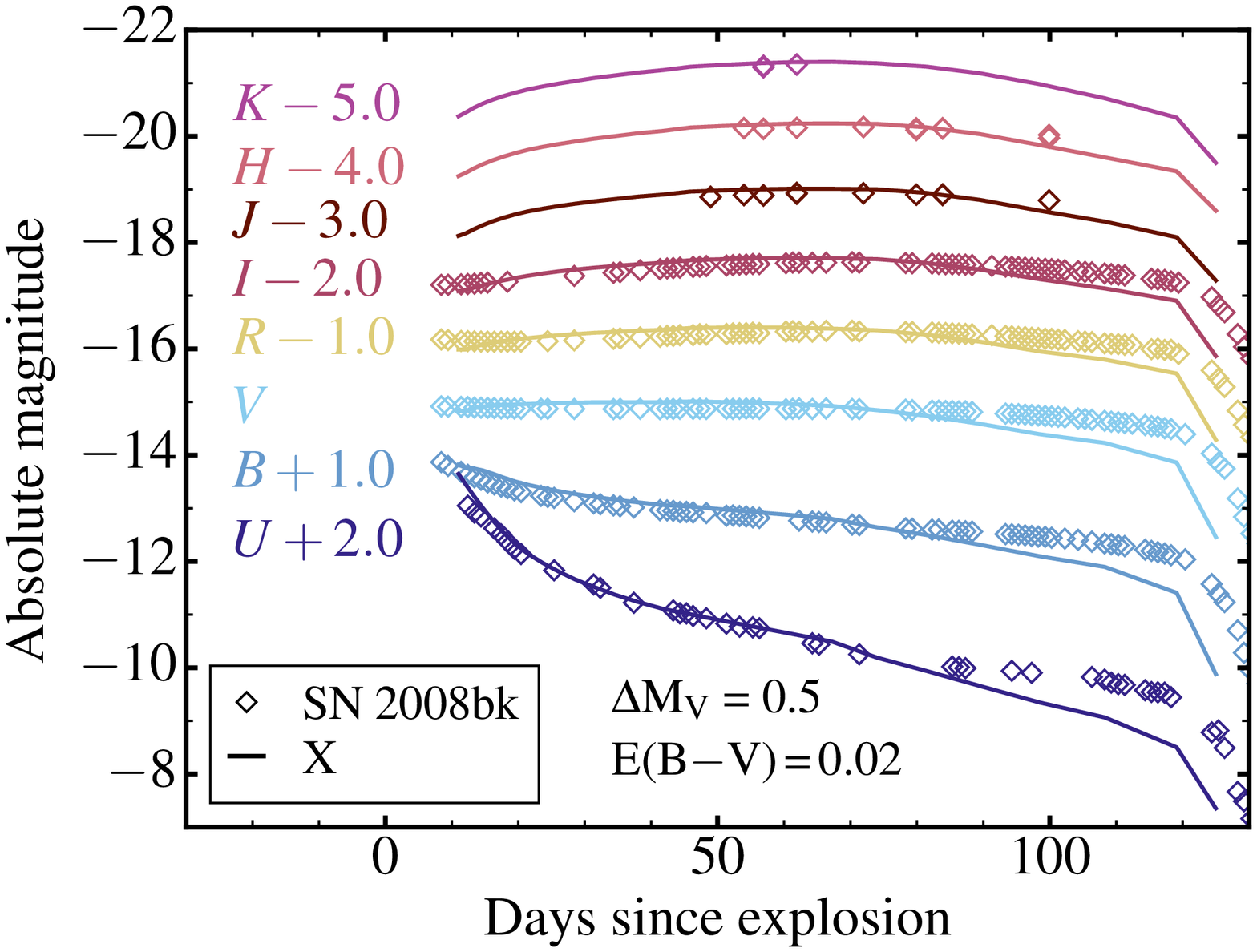}
  \includegraphics[width=0.48\textwidth]{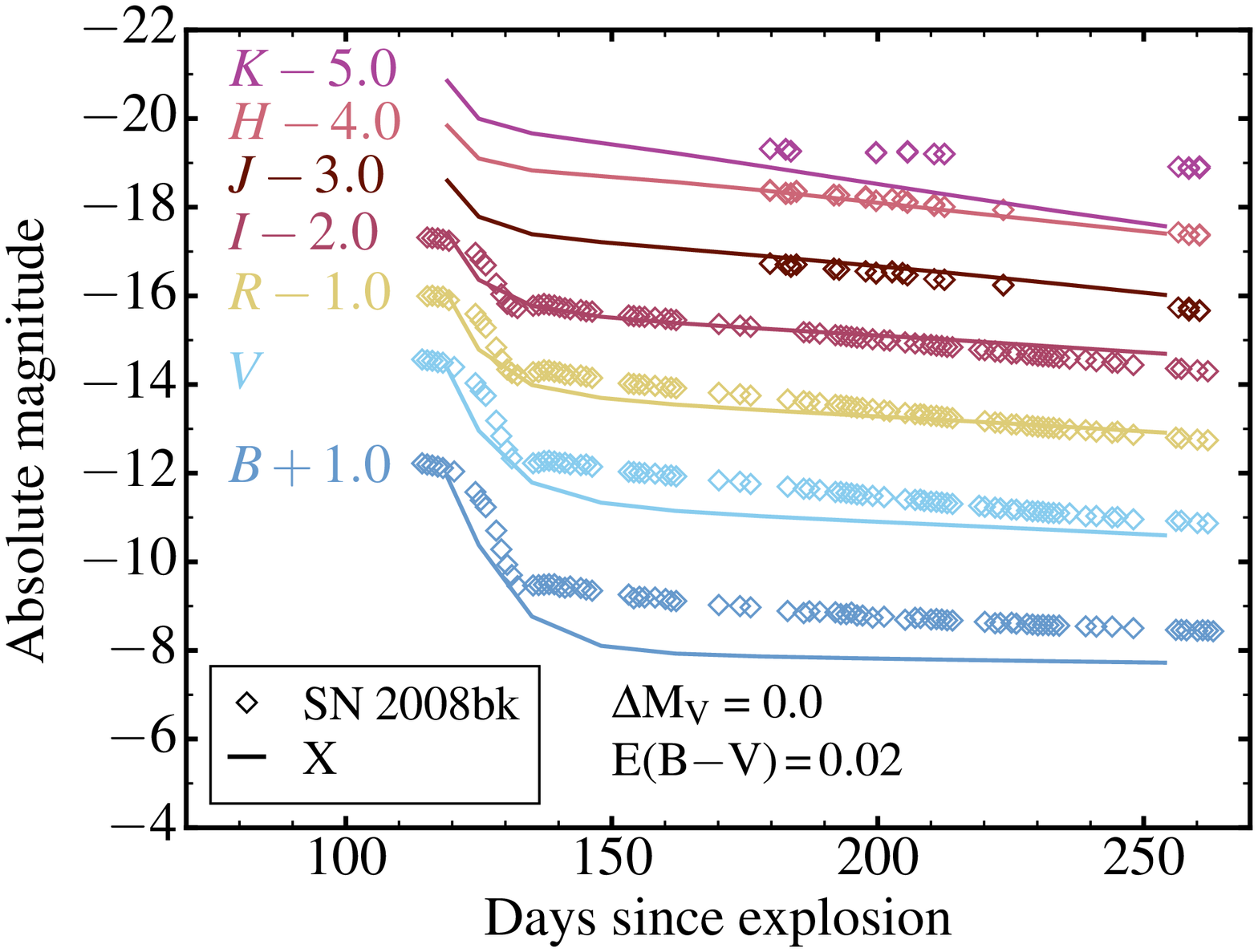}
  \caption{
  Comparison of optical and near-IR light curves for SN 2008bk 
  (corrected for reddening and distance) and model X
  during the plateau phase (left) and the nebular phase (right).
  At early times model X is somewhat over luminous, and we have adjusted all magnitudes 
by $\Delta M_V = M_{V, \rm obs}- M_{V, \rm model}=$\,0.5\,mag to facilitate the
comparison with SN 2008bk. During the nebula phase the luminosity is set primarily by
the original 56Ni mass, and no adjustment was necessary (See Section~\ref{sect_data} and
discussion in Section~\ref{sect_phot}).
  \label{fig_lc_X_08bk}
  }
\end{figure*}

\begin{figure}
  \includegraphics[width=0.5\textwidth]{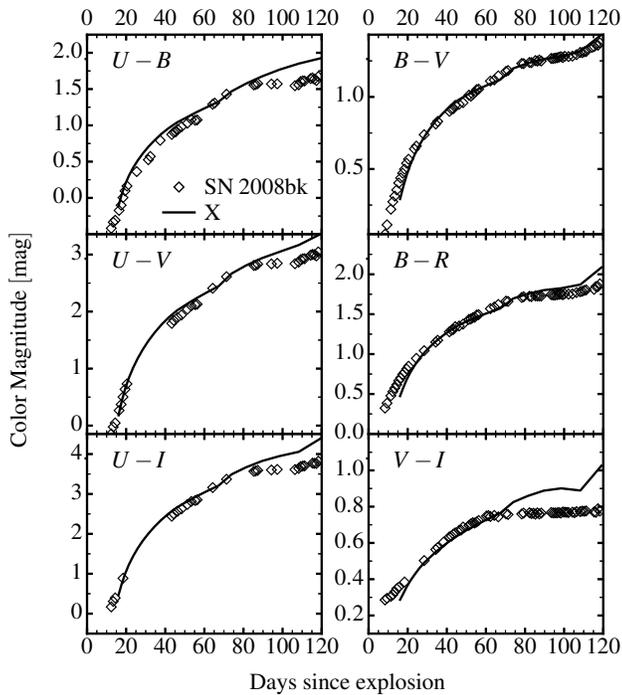}
    \caption{Color evolution for our best-match model X and the de-reddened observations
    of SN\,2008bk. See Section~\ref{sect_data} for details on the observational data.
    \label{fig_X_color}}
\end{figure}

\subsection{Radiative-transfer modelling}
\label{sect_cmfgen}

   When the ejecta reach homologous expansion, we remap each model into
   the nLTE time-dependent radiative-transfer code \cmfgen\
   \citep{HM98,DH05a,DH08, HD12,dessart_etal_13}
   to compute the subsequent evolution of the gas and the radiation until nebular times.
   The code computes the gas and radiation properties by solving iteratively the
   statistical equilibrium equations, the gas-energy equation, and the first two moments of the radiative
   transfer equation.
   Time-dependent terms are included in all equations.
   Non-thermal processes and thermal heating associated with radioactive decay are included.
   For each converged model, the code produces a spectrum with a sub-\AA\ resolution that covers
   from the far-UV to the far-IR and from which one can extract the bolometric luminosity and various
   photometric magnitudes. It is thus the same tool that produces multi-band light curves and spectra.
   We can directly compare the multi-band light curves computed by \cmfgen\ with the observed ones,
   constraining the reddening from the colour and spectral information. 
   
    All time sequences with \cmfgen\ are started at $\sim$\,11\,d. The entire progenitor H-rich envelope
    is in homologous expansion by then, but not the slow moving He-core material.
    So, by enforcing homologous expansion, we modify slightly the velocity of the He-core material,
    i.e. in regions moving with $<$700-800\,\kms\ (see Fig.~\ref{fig_X_comp}).
   Homology requires that we reset the time to $R/V$, which causes a shift to an earlier time by $\sim$\,0.5\,d. 
    This new time differs from the time elapsed since explosive nucleosynthesis took place, making the \iso{56}Ni 
    and \iso{56}Co masses incompatible. In practice, the \iso{56}Co mass is overestimated by $\sim$\,10\%,
    so the effect is minor. For consistency, we quote for each model an initial \iso{56}Ni masse inferred 
    from the ejecta \iso{56}Co mass at nebular times.

    The \cmfgen\ simulations use 100 grid points, which are placed to adequately resolve
the variations in optical depth. This is an asset over a radiation hydrodynamics code
like \v1d, which uses a grid tied to the Lagrangian mass. With this choice, our \cmfgen\ simulations
resolve well the (moving) recombination fronts associated with H and He (see \citealt{DH10}).

    The numerical setup is comparable to that of \citet{dessart_etal_13}. We use the same model atoms,
    with updates to the atomic data (in particular for Fe and Co) as described in \citet{D14}.
    We treat the following ions:  H\one, He\one-\two, C\one-\four, N\one-\three, O\one-\five,
     Ne\one-\three, Na\one, Mg\one-\three, Si\one-\four, S\one-\four, Ar\one-\three, K\one, Ca\one-\four,
     Sc\one-\three, Ti\two-\three, Cr\two-\four, Fe\one-\six, Co\two-\six, and Ni\two-\six.

    We include only the radioactive decay from \iso{56}Ni and \iso{56}Co -- no other unstable
     isotope is considered in the present calculations.

     In one instance, we recompute the model X at one time step with the addition of Ba\two\ in order to test the
     impact of that ion on the spectrum at the end of the photospheric phase (Section~\ref{sect_spec}).

    We show the bolometric light curve for all models in Fig.~\ref{lc_lbol}. The maximum ``plateau" luminosity is in the range
    $\sim$\,10$^{41.5}-10^{42.0}$\,\ergs, and the plateau duration is in the range $\sim$\,120--140\,d.
    Each model will be discussed in turn in the following sections,
    starting with a detailed presentation of model X, our best-match model to the observations
    of SN\,2008bk. A summary of the results is presented in Table~\ref{tab:res}.


\section{Properties of our best-match model to the observations of SN\,2008\lowercase{bk}}
\label{sect_X}

   Model X stems from a 12\,\msun\ main-sequence star evolved at solar metallicity and dying with a 
   pre-SN radius of 502\,\rsun. The resulting SN ejecta is 8.29\,\msun, with a kinetic energy of 2.5$\times$10$^{50}$\,erg
and 0.0086\,\msun\ of \iso{56}Ni. These parameters do not yield a perfect match to observations
of SN\,2008bk and therefore do not correspond exactly
to the progenitor and the ejecta of SN\,2008bk, but they produce a reasonable match to the observations.
In this section, we present the photometric and spectroscopic properties of model X. We also 
discuss the possible origin of the discrepancies and how they may be reduced to provide
a better match.

\subsection{Photometric properties}
\label{sect_phot}

Figure~\ref{fig_lc_X_08bk} compares the optical ($UBVRI$) and near-IR ($JHK$) light curves
of model X with the observations of SN\,2008bk. A global magnitude offset (equivalent to a shift in distance modulus)
has been applied because model X is somewhat too bright during the photospheric phase
(the offset is of 0.5\,mag, equivalent to a $\lesssim$\,60\% overestimate in luminosity),
and somewhat too faint during the nebular phase.

The overestimate of the plateau brightness suggests that model X is somewhat too energetic
and/or its progenitor radius is somewhat too large (Section~\ref{sect_rad}).
The Cepheid-based distance to the galaxy hosting SN\,2008bk is unlikely to be a sizeable source of
error, but the adopted reddening may be underestimated.
If we adopt a reddening $E(B-V)=$\,0.1\,mag, the match to the colour evolution is somewhat degraded
but the absolute offset is then reduced from 0.5 to 0.2\,mag.
At nebular times, model X is under-luminous in the $B$ and $V$ bands, but it matches satisfactorily
the redder bands $RIJH$ --- most of the flux falls in the range 5000-10000\,\AA\ at nebular times.
As we discuss in Section~\ref{sect_mixing}, the mixing has a large impact on the SN optical colors at
nebular times. However, given the reasonable match to the filter bands where most of the flux falls (within
the range 5000-10000\,\AA), the 0.0086\,\msun\ mass of \iso{56}Ni in model X is satisfactory
(given the adopted distance/reddening).

Leaving the color mismatch aside, Fig.~\ref{fig_lc_X_08bk} shows that the multi-band light curve
of model X matches adequately the multi-band light curve evolution from the $U$-band to the near-IR.
The match to the steep drop in the $U$ band suggests the small progenitor radius of 502\,\rsun\ is roughly
adequate, something that arises in the \mesa\ simulation of the progenitor star from the adoption of an enhanced
mixing-length parameter for convection \citep{dessart_etal_13}. As we discuss in Section~\ref{sect_rad},
a change of 10-20\% in progenitor radius has a visible impact on the colour evolution.

During the photospheric phase, the SN radiation is comparable to a blackbody modified by the effect
of scattering, atmospheric extension, and line-blanketing, and it is mostly influenced by the global parameters
of the progenitor star and explosion, namely $M_\star$, $R_\star$, and $E_{\rm kin}$.
These aspects seem to be properly modelled here.
However, at nebular times (right panel of Fig.~\ref{fig_lc_X_08bk}), the spectrum is less directly
connected to  $M_\star$, $R_\star$, and $E_{\rm kin}$. Instead, it becomes primarily
influenced by heating from radioactive decay and cooling from line emission. The spectrum is then
more directly sensitive to chemical composition and mixing.
It is therefore not surprising that model X can match better the photospheric phase than the nebular
phase of SN\,2008bk (see Section~\ref{sect_mixing} for discussion).

\begin{figure*}
  \includegraphics[width=0.48\textwidth]{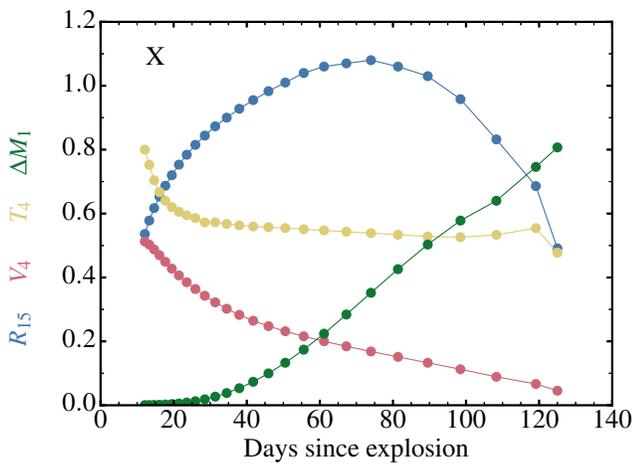}
  \includegraphics[width=0.48\textwidth]{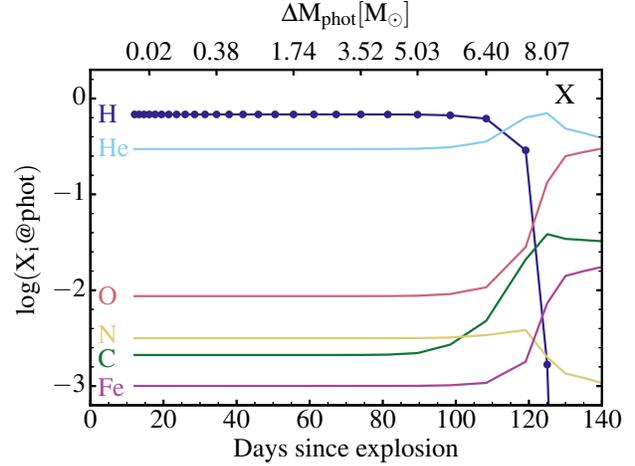}
  \caption{
  Evolution of the properties at the photosphere of our best-match model X.
  In the left panel, we show the evolution of the photospheric radius $R_{15}$ 
  ($R_{15}=R_{\rm phot}/10^{15}$\,cm), the photospheric  velocity $V_4$ 
  ($V_4=V_{\rm phot}/10^4$\,\kms), the photospheric temperature $T_4$
  ($T_4=T_{\rm phot}/10^4$\,K), and the mass above the photosphere $\Delta
  M_1$ ($\Delta M_1=\Delta M_{\rm phot}/10\,$\msun) until 140\,d after
  explosion.  The dots correspond to the actual times at which the
  radiation-transfer simulations were performed.
  In the right panel, we show the evolution of the composition at the photosphere for a few species.
\label{fig_X_phot}
  }
\end{figure*}

The colour evolution of model X during the photospheric phase is also in good agreement with the observations
of SN\,2008bk (Fig.~\ref{fig_X_color}).
In Type II SNe, the color evolution is monotonic and follows the progressive decrease
of the photospheric temperature (Fig.~\ref{fig_X_phot}; see also Table~\ref{tab_X_phot}),
from the high temperature conditions at shock breakout until the
recombination phase where the photospheric temperature is set by the recombination temperature of hydrogen.
This phase starts in model X at 20--30\,d and is associated with a flattening of the evolution of $T_{\rm phot}$,
which levels off at $\sim$\,5500\,K --- $R_{\rm phot}$ is then of the order of 10$^{15}$\,cm.
This is in contrast to \iso{56}Ni powered SNe (e.g., SNe Ia, Ib, and Ic), in which decay heating leads
to a non-monotonic evolution of the colour during the photospheric phase.
A $\lesssim$\,0.1\,mag colour offset appears at the end of the high-brightness phase (the model is too red),
in particular for ($V-I$) --- this is also visible in Fig.~\ref{fig_lc_X_08bk}. This small offset
may be related to the adopted mixing between the He-core
and the H-rich envelope (which is treated crudely in our approach), an overestimate
of line blanketing (perhaps connected to a problem with the atomic data, e.g., with Fe\one; see next section), or
a problem with the data reduction of the $I$-band photometry (Pignata, priv. communication).

Compared to models of standard SNe II-P computed in the past with the same approach,
model X is underluminous, with a typical bolometric luminosity of $\sim$\,1.1$\times$10$^8$\,\lsun.
For comparison, the 1.2$\times$10$^{51}$\,erg s15 model of \citet{DH11} had a plateau luminosity
of $\sim$\,6$\times$10$^8$\,\lsun\ and the more compact progenitor model MLT3 of \citet{dessart_etal_13}
had a plateau luminosity of $\sim$\,4$\times$10$^8$\,\lsun.
Both had the same photospheric temperature of $\sim$5500\,K at the recombination epoch.
Instead,  model X differs from these more energetic explosions by a much reduced photospheric radius.
Model X has a radius of 10$^{15}$\,cm, a factor of 2--3 times smaller than models s15 and MLT3.

Another interesting feature, shared by models X and the higher energy variants s15 and MLT3, is that at 50\,d after explosion,
the photosphere has not receded by more than 1\,\msun\ below the progenitor surface
(see also, e.g., \citealt{FA77}).
This implies that when intrinsically
polarised continuum radiation is observed soon after explosion in a SN II-P \citep{leonard_12aw_12},
the outermost ejecta layers have to be asymmetric.

The photospheric phase (i.e., when the ejecta Rosseland-mean or electron-scattering
optical depth exceeds 2/3), or the high-brightness phase, correspond to epochs when the spectrum
forms in the H-rich ejecta layers (see, e.g., \citealt{KW09}, \citealt{DH11}).
Both the optical depth and the brightness plummet when the photosphere recedes into the layers rich in helium and
intermediate-mass elements (regions that  were formally part of the highly-bound compact
He core), at velocities below $\sim$\,1000\,\kms\ in model X (see details in Table~\ref{tab_X_phot}
and illustrations in Fig.~\ref{fig_X_phot}).

\begin{table}
\begin{center}
\caption{
Evolution during the photospheric phase of the total ejecta electron-scattering optical depth, and of the radius, velocity, temperature, and overlying mass
at the electron-scattering photosphere (using the Rosseland-mean opacity would shift the photosphere slightly outwards
in radius/velocity space). We also include the (observer's frame) bolometric luminosity.
Numbers in parenthesis correspond to powers of ten.
\label{tab_X_phot}
}
\begin{tabular}{l@{\hspace{2mm}}c@{\hspace{2mm}}c@{\hspace{2mm}}
c@{\hspace{2mm}}c@{\hspace{2mm}}c@{\hspace{2mm}}c@{\hspace{2mm}}}
\hline
Age &   $\tau_{\rm base, es}$ & $R_{\rm phot}$ &   $V_{\rm phot}$ &   $T_{\rm phot}$ &   $\Delta M_{\rm phot}$  & $L_{\rm bol}$ \\
\hline
[d]      &    & [cm]                  &  [\kms]                   &   [K]                      & [\msun]   & [10$^8$\,\lsun] \\
\hline
     12.10     &      26324    &         5.36(14)    &       5124   &  	       8000   	 &       5.66(-3)    &            1.25   \\
     13.31     &      21746    &         5.78(14)    &       5028   &  	       7526    	 &       7.45(-3)    &            1.18   \\
     14.64     &      17968    &         6.17(14)    &       4878   &  	       7040    	 &       1.06(-2)    &            1.15   \\
     16.10     &      14841    &         6.53(14)    &       4694   &  	       6671    	 &       1.55(-2)    &            1.12 \\
     17.71     &      12231    &         6.87(14)    &       4489   &  	       6403    	 &       2.33(-2)    &            1.10   \\
     19.48     &      10067    &         7.20(14)    &       4276   &  	       6203    	 &       3.56(-2)    &            1.09     \\
     21.43     &       8274    &         7.53(14)    &       4064   &  	       6055    	 &       5.48(-2)    &            1.09   \\
     23.57     &       6803    &         7.84(14)    &       3851   &  	       5946    	 &       8.42(-2)    &            1.09   \\
     25.93     &       5584    &         8.15(14)    &       3638   &  	       5856    	 &       1.26(-1)    &            1.09   \\
     28.52     &       4568    &         8.44(14)    &       3427   &  	       5723    	 &       1.86(-1)    &            1.09   \\
     31.37     &       3711    &         8.73(14)    &       3221   &  	       5726    	 &       2.69(-1)    &            1.10   \\
     34.51     &       2973    &         9.00(14)    &       3020   &  	       5675    	 &       3.82(-1)    &            1.10   \\
     37.96     &       2335    &         9.28(14)    &       2828   &  	       5633    	 &       5.33(-1)    &            1.12   \\
     41.76     &       1818    &         9.55(14)    &       2646   &  	       5597    	 &       7.35(-1)    &            1.13   \\
     45.94     &       1416    &         9.83(14)    &       2476   &  	       5573    	 &       9.97(-1)    &            1.16   \\
     50.53     &       1100    &         1.01(15)    &       2316   &  	       5546    	 &       1.33        &        1.16	     \\
     55.58     &        857    &         1.04(15)    &       2158   &  	       5508    	 &       1.74        &        1.16	     \\
     61.14     &        671    &         1.06(15)    &       2002   &  	       5472    	 &       2.24        &        1.16	     \\
     67.25     &        521    &         1.07(15)    &       1845   &  	       5431    	 &       2.84        &        1.13	     \\
     73.98     &        395    &         1.08(15)    &       1683   &  	       5388    	 &       3.52        &        1.09	     \\
     81.38     &        288    &         1.06(15)    &       1514   &  	       5338    	 &       4.26        &        1.01	     \\
     89.52     &        195    &         1.03(15)    &       1330   &  	       5281    	 &       5.03        &       0.885	     \\
     98.47     &        114    &         9.58(14)    &       1126   &  	       5260    	 &       5.78        &       0.737	     \\
    108.30     &         45    &         8.32(14)    &        889   &  	       5332    	 &       6.40        &       0.619	     \\
    119.10     &         4.1   &          6.86(14)   &         666  &   	        5544          &      7.46         &           0.470	            \\
    125.00     &         1.2   &          4.91(14)   &         455  &   	        4775          &     8.07          &   0.173	               \\
\hline
\end{tabular}
\end{center}
\end{table}

\subsection{Spectroscopic properties}
\label{sect_spec}

\begin{figure*}
\includegraphics[width=0.98\textwidth]{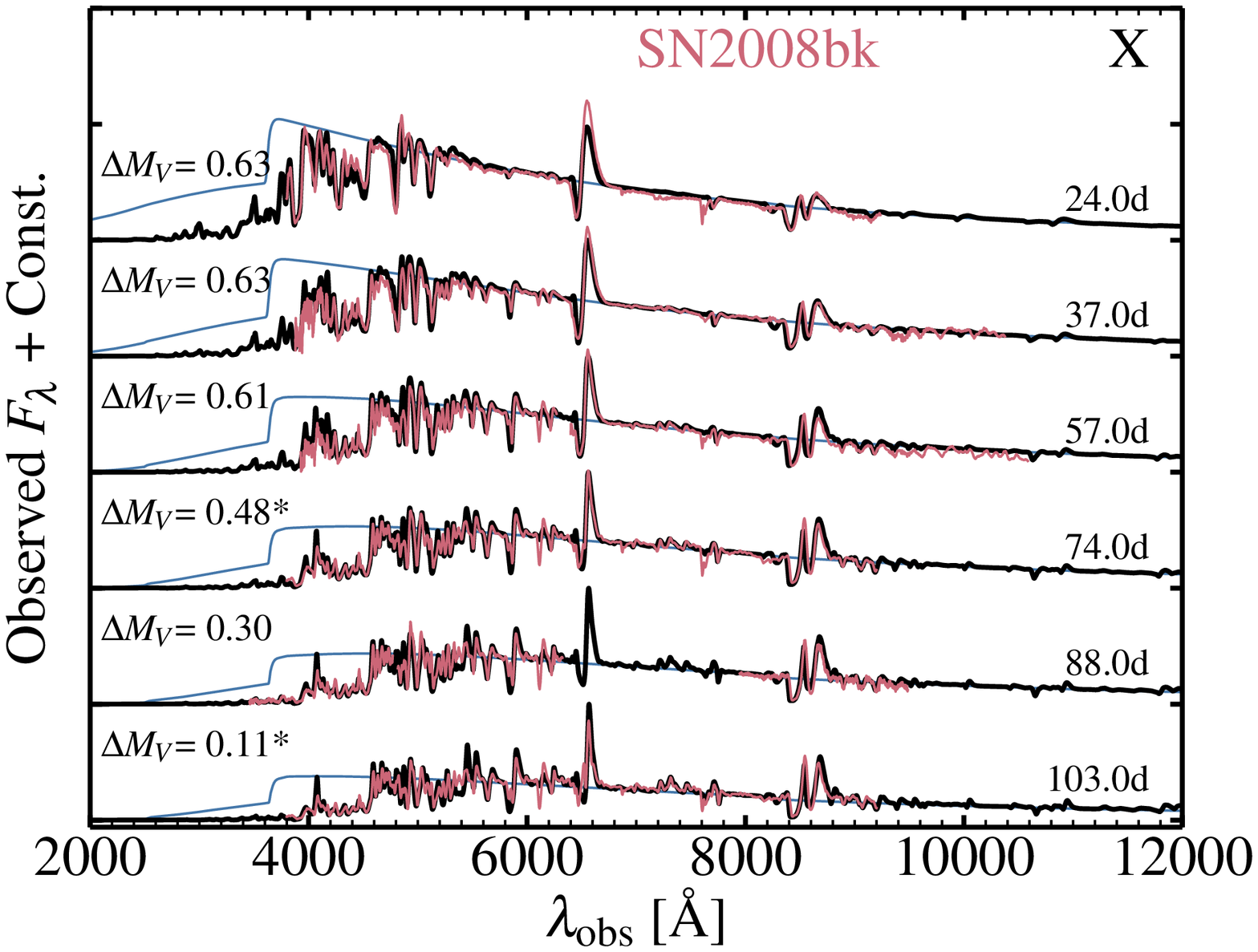}
  \caption{Comparison of synthetic spectra for our best-match model X (black: full spectrum;
  blue: continuum only) with the
  observations of  SN\,2008bk at multiple epoch during the photospheric phase.
  The model spectra have been reddened, red-shifted, and distance-scaled (see Section~\ref{sect_data}).
  To facilitate the comparison of spectral features, a global flux shift has been applied to each spectrum
  (see the corresponding magnitude offset $\Delta M_V = M_{V, \rm obs}- M_{V, \rm model}$).
  In addition, when this label has a star superscript, a scaling of the relative
  flux distribution has been applied to match the model flux (see Section~\ref{sect_data} for discussion).
  The poor match to the 6150\,\AA\ feature is caused by the neglect of Ba\two\ lines in these spectral calculations
  (but see Section~\ref{sect_ba2}).
  [See text for discussion.]
\label{fig_X_spec_obs}}
\end{figure*}

Figure~\ref{fig_X_spec_obs} shows a comparison of the spectroscopic observations of SN\,2008bk
with model X at multiple epochs during the photospheric phase. The nebular phase will be discussed
in Section~\ref{sect_mixing} and in a separate study.
Here, we have corrected the model for reddening, redshift, and distance to directly compare
to the observations. As before for the multi-band light curves at early times, we apply a global scaling of
the model that corresponds to a magnitude offset of $\sim$\,0.5\,mag (the model is 70\% too luminous for our adopted
distance and reddening --- see discussion in the previous section).
Figure~\ref{fig_X_ladder} shows the bound-bound transitions that influence the photospheric
spectra at 23.6 and 108.3\,d.

While the photometric observations of SN\,2008bk started within a week of explosion, the spectroscopic
monitoring started only three weeks after explosion. The SN has entered the recombination phase (see Fig.~\ref{fig_X_phot}),
the spectral energy distribution is already quite red, and we see clear signs of line blanketing (Fe\two\ and Ti\two) in the optical.
This is made more obvious by comparing the synthetic spectrum (black) with the continuum flux from the model (blue; the
continuum flux is computed, in a post-processing step, by including only the bound-free and free-free
processes in the formal solution of the radiative transfer equation).
Line blanketing suppresses strongly the flux shortward of 5000\,\AA, although one clearly sees that the
continuum flux is not strong below the Balmer edge, as expected for a cool photosphere
(Table~\ref{tab_X_phot} and Fig.~\ref{tab_X_phot}).
Similarly, He\one\ lines have already vanished and Na\one\,D (rather than He\one\,5875\,\AA)
is already present in the first spectrum.
We see H\one\ Balmer lines (and in particular H$\alpha$) together with Ca\two\,H\&K, the Ca\two\ near-IR triplet,
O\one\,7773.4\,\AA, and a trace of Si\two\,6355\,\AA.

The spectrum evolves slowly through the photospheric phase. Line blanketing strengthens 
as the photospheric temperature progressively drops and the line profiles 
become increasingly narrow as the photosphere recedes
into deeper/slower ejecta layers (Table~\ref{tab_X_phot}).
The photospheric velocity of model X is $\gtrsim$\,2000\,\kms\ halfway through the plateau phase,
about 10-20\% too large for SN\,2008bk, but 50\% lower than for a standard SN II-P like SN\,1999em
\citep{DH06,utrobin_99em_07,pignata_08bk_13,bersten_11_2p,dessart_etal_13}.
As we approach the end of the plateau, the ionization level at the photosphere decreases so ions with
a lower ionization potential start contributing. We see the strengthening of lines from
Fe\one\ and Sc\two, as well as from Fe\two\ in the red part of the optical.

 \begin{figure}
  \includegraphics[width=0.48\textwidth]{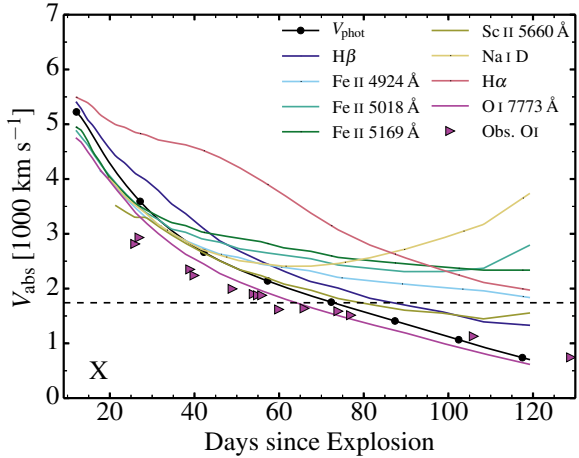}
\caption{
Evolution of the velocity at maximum absorption for the bound-bound transitions at
H$\beta$, Fe\two\,4924, Fe\two\,5018, Fe\two\,5169, Sc\two\,5660, Na\one\,D, H$\alpha$,
and O\one\,7773\,\AA.
The dashed line corresponds to the quantity $V_{\rm m} \equiv \sqrt{2E_{\rm kin}/M_{\rm e}}=$\,1741\,\kms,
and the line with dots corresponds to the photospheric velocity (calculated using the Rosseland-mean opacity).
The filled triangles correspond to the observed velocity at maximum absorption for O\one\,7773\,\AA\ in SN\,2008bk.
\label{fig_X_vabs}
}
\end{figure}

\begin{figure}
\includegraphics[width=0.48\textwidth]{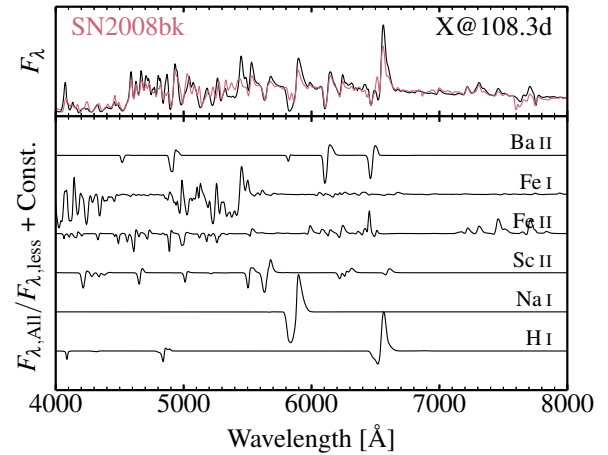}
  \caption{
  Comparison between the observations of SN\,2008bk taken on the 1st of July 2008 (corresponding to
  an inferred post-explosion time of 103.0\,d) and a calculation for model X at 108.3\,d that includes Ba\two.
  In low-luminosity SNe II-P like 2008bk, the structure in H$\alpha$ at the end of the plateau phase is caused
  by overlap with Ba\two\,6496.9\,\AA\ (not perfectly fitted here), a conclusion that is reinforced by the good
  match to the isolated Ba\two\,6141.7\,\AA\ line (compare with Fig.~\ref{fig_X_spec_obs} for the prediction
  with the model that does not include Ba\two).
    \label{fig_ba2}}
\end{figure}

 Figure~\ref{fig_X_vabs} illustrates the evolution of the velocity at maximum absorption in strong optical
 lines of  H\one, Fe\two, Sc\two, Na\one, and O\one. We see that the line that matches best the evolution
 of the photosphere (shown in black)
 is O\one\,7773.4\,\AA. Fe\two\,5169\,\AA, which is often used for estimating $V_{\rm phot}$, tends to overestimate
 it in the second half of the plateau, while it underestimates it at early times (see discussion in \citealt{DH05b}).
 Compared to SN\,2008bk, the expansion rate of the model X ejecta is too large by 10--20\%. The overestimate
 of  the velocity at maximum absorption by model X is in the same sense as for the luminosity;
 our model X is slightly over-energetic for its mass and radius.

\subsection{Ba\two\ lines and the structure seen in H$\alpha$}
\label{sect_ba2}

The spectral morphology is markedly different from standard SNe II; line profiles are narrower and suffer much
less overlap with neighbouring features.
At the end of the photospheric phase the H$\alpha$ profile in SN\,2008bk
shows a complex structure, which is absent in model X (Fig.~\ref{fig_X_spec_obs}).
This may be a signature of asphericity in the inner ejecta, which our 1-D approach cannot capture.
But a simpler alternative is the contamination from Ba\two\ lines. This seems likely since we strongly underestimate
the strength of the feature at 6150\,\AA, which may stem primarily from Ba\two\,6141.7\,\AA.
Ba was not included in any time sequence here because it is an s-process element not treated in our \mesa\ simulations.

Most low-luminosity SNe II-P exhibit a structured H$\alpha$ profile at the end
of the photospheric phase, while standard SNe II-P tend not to (see Fig.~6 of \citealt{roy_08in_11}).
To investigate whether Ba\two\ is at the origin of this complex structure, we recomputed model X at one
time step with Ba\two\ included. Because we did not have Ba in the time-dependent \cmfgen\ model, we used Sc\two\
to set the Ba\two\ departure coefficients at the previous time step. We also initialise the Ba abundance in the
H-rich envelope and in the inner ejecta to the same values as those obtained in the detailed \kepler\ model s15iso presented
in \citet{DH11} --- our adopted Ba mass fraction in the H-rich envelope is 1.32$\times$10$^{-8}$.
As shown in Fig.~\ref{fig_ba2},  with the inclusion of Ba\two\ in the model atom, we now reproduce
the observed feature at 6150\,\AA, and we also obtain a much more structured H$\alpha$ line profile,
although the model profile does not match exactly the observations.
We also predict a weaker Ba\two\ line overlapping with Na\one\,D: this line has a double dip
in the observations but the model shows just one broad absorption, perhaps because the model overestimates
the range of velocities over which Na\one\,D forms.
The expansion rate in model X is slightly larger than needed for SN\,2008bk (see also Fig.~\ref{fig_X_vabs}),
so the broader lines of the model tend to overestimate line overlap and erase the structure in the H$\alpha$ region.

We surmise that the complex appearance of the H$\alpha$ region at the end of the photospheric
phase in low-luminosity SNe II is due to Ba\two\,6496.9\,\AA.
In standard SNe II the higher expansion rate at the base of the H-rich envelope causes the contributions
of  Ba\two\,6496.9\,\AA\ and H$\alpha$ to merge into a single spectral feature.
In addition to the influence of Ba\two, interaction with the progenitor RSG wind
(ignored in our model) may influence at such late times the H$\alpha$ and H$\beta$ line profile morphology
through the formation of a `high-velocity' absorption notch \citep{chugai_hv_07}.
It is however unclear whether the effect would apply for our model X since its low mass RSG progenitor
should have a low wind mass loss rate.

\subsection{Additional remarks}

Leaving aside the discrepancies associated with the neglect of Ba\two, Fig.~\ref{fig_X_spec_obs} shows that model X
reproduces closely the observed spectra of SN\,2008bk ---
no tuning is done to the composition either initially (we just adopt the composition
of the \mesa\ model evolved at solar metallicity) or in the course of the \cmfgen\ time sequence.
Our model seems to overestimate the absorption at 5400\,\AA\ associated with Fe\one\ lines 
(see appendix), perhaps because of problems
with the atomic data (the strength of Fe\two\ lines seems adequate). The features around 9000\,\AA--1\,$\mu$m
seen in the observations of SN\,2008bk at 35 and 55\,d are not predicted by our models --- these are
perhaps observational artefacts (e.g., second order contamination when using a dichroic).


\begin{figure}
  \includegraphics[width=0.48\textwidth]{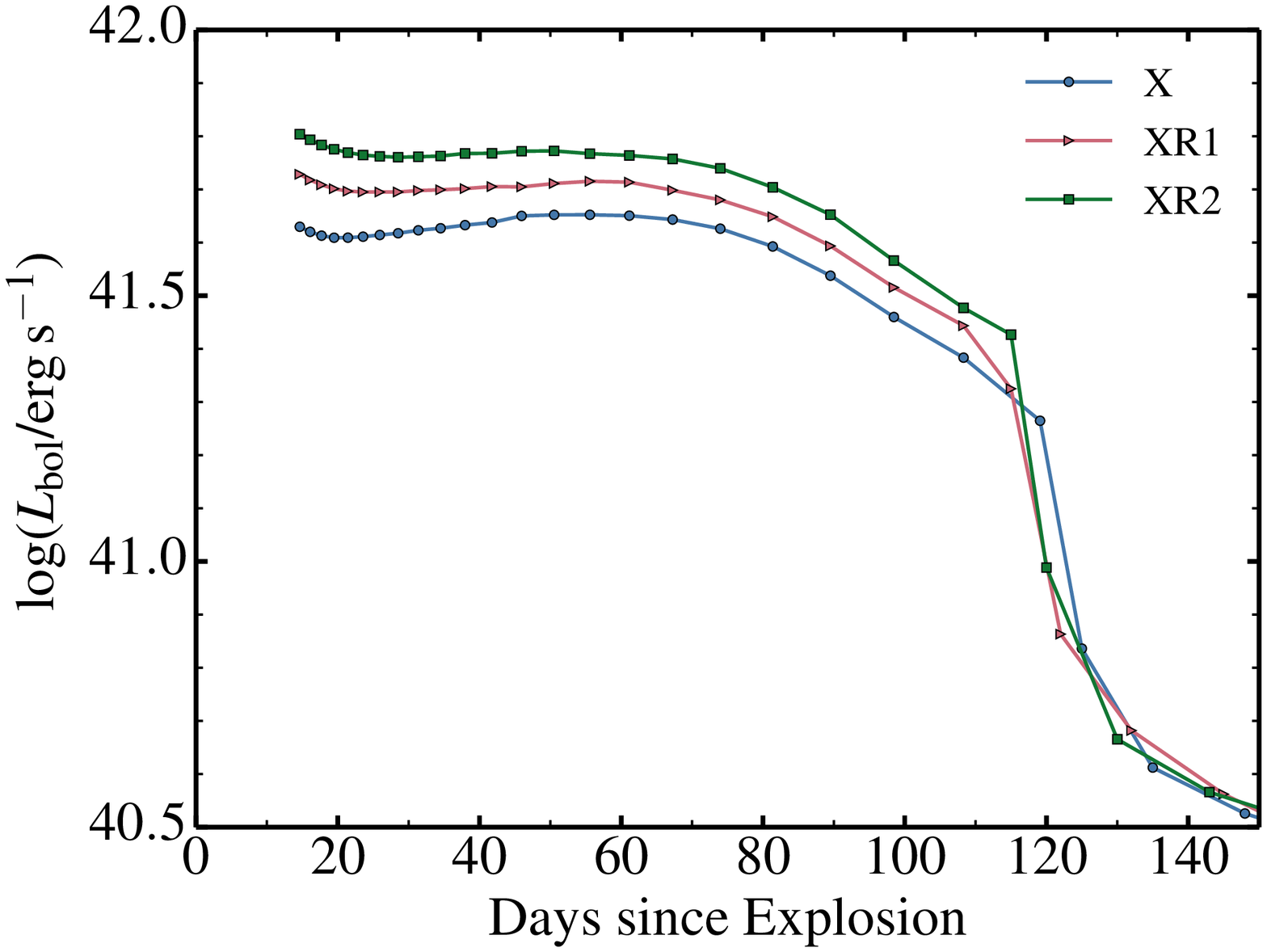}
  \includegraphics[width=0.48\textwidth]{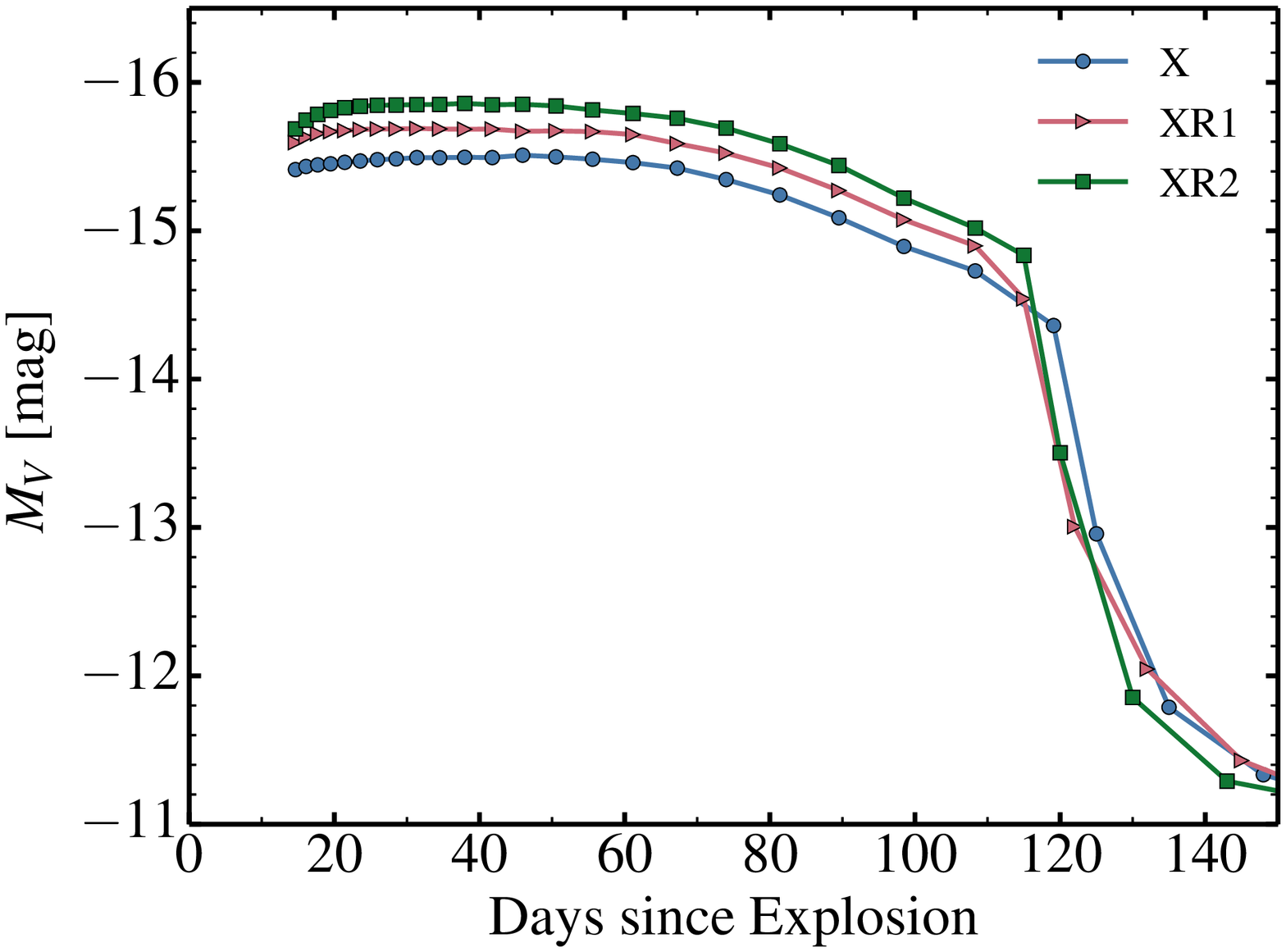}
  \includegraphics[width=0.48\textwidth]{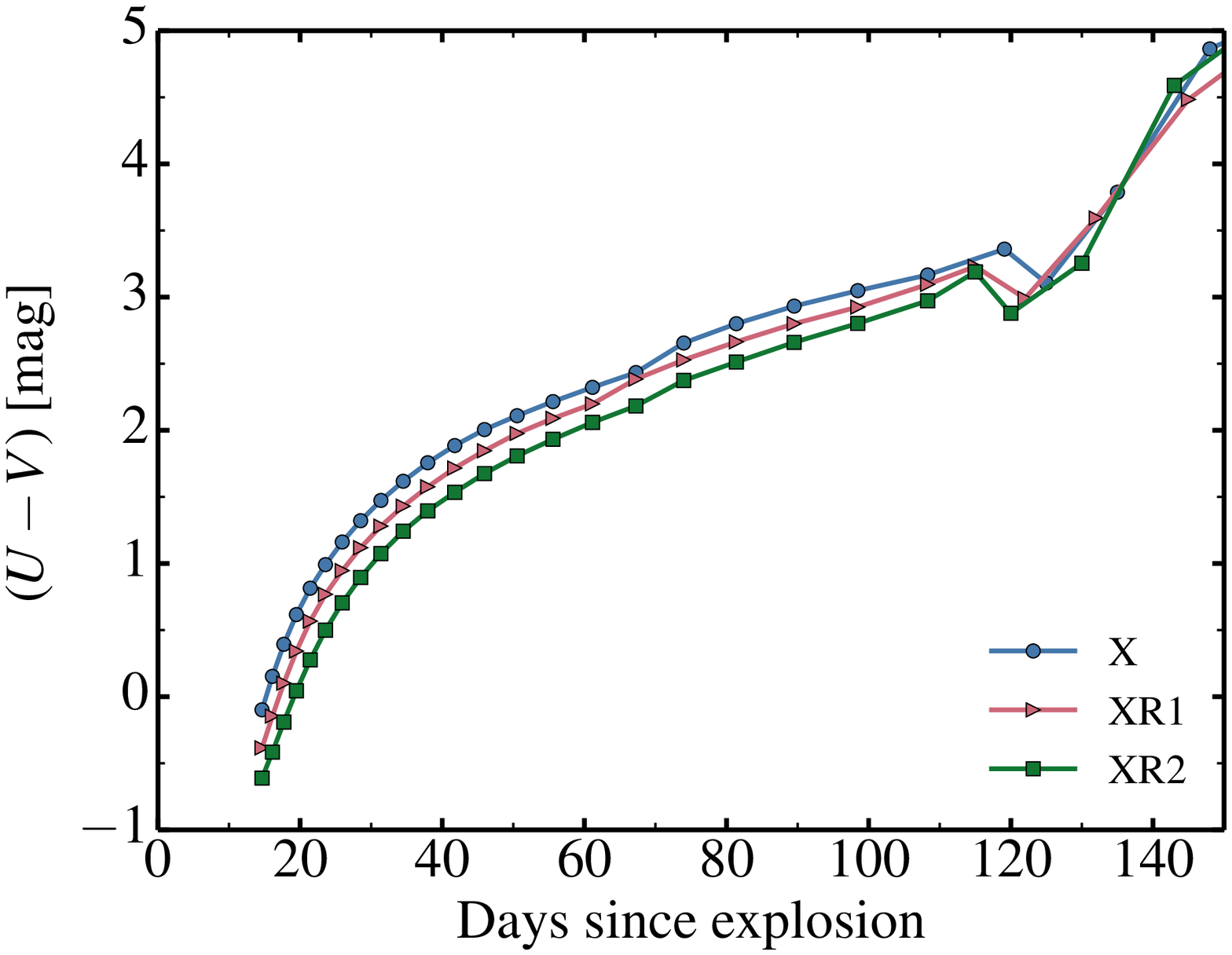}
    \caption{
    Bolometric luminosity (top), $V$-band magnitude (middle) and $(U-V)$ colour (bottom) for
    models X, XR1, and XR2, which differ primarily in the surface radius of their RSG progenitor star.
\label{fig_lc_rad}}
\end{figure}

\begin{figure}
  \includegraphics[width=0.48\textwidth]{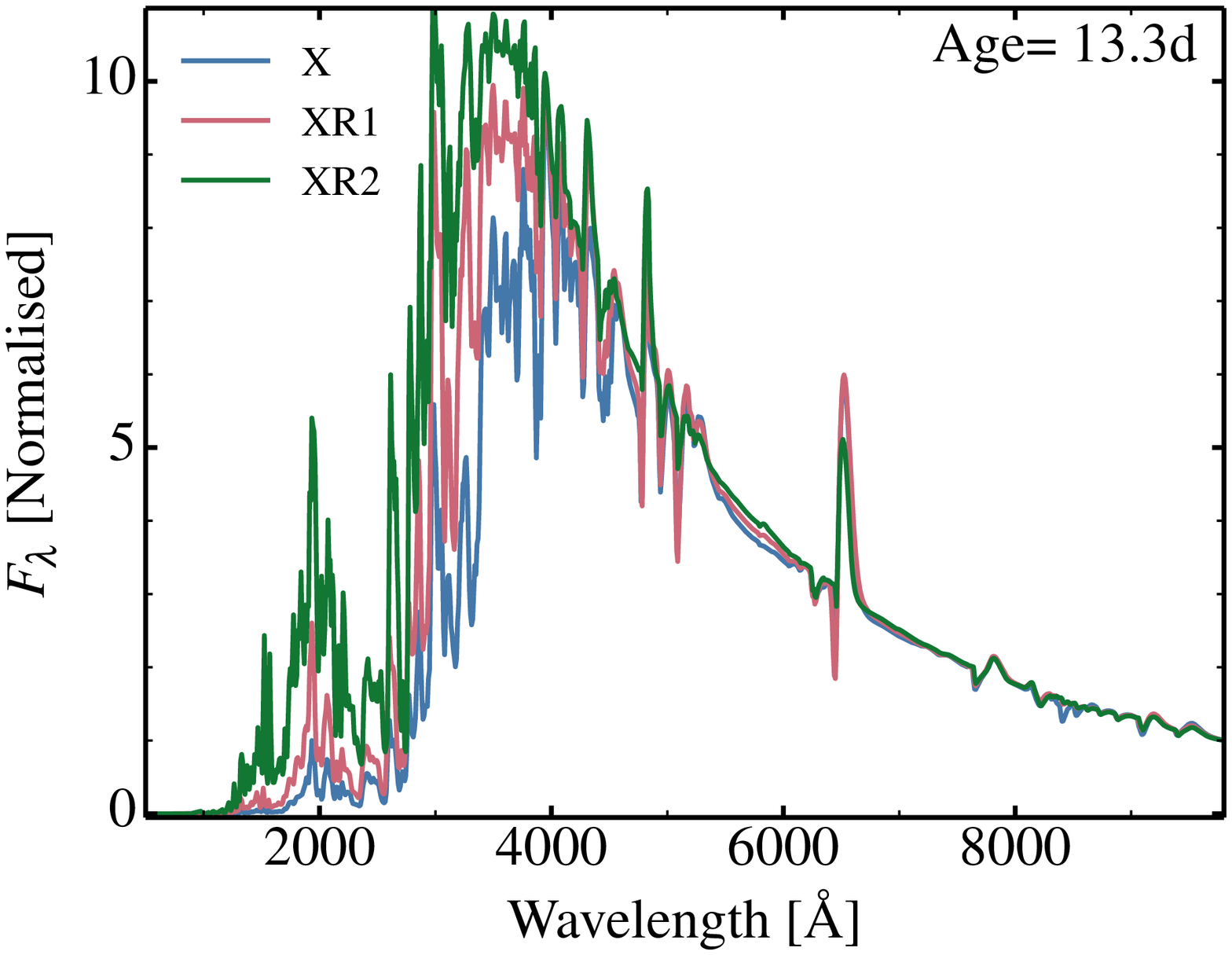}
  \includegraphics[width=0.48\textwidth]{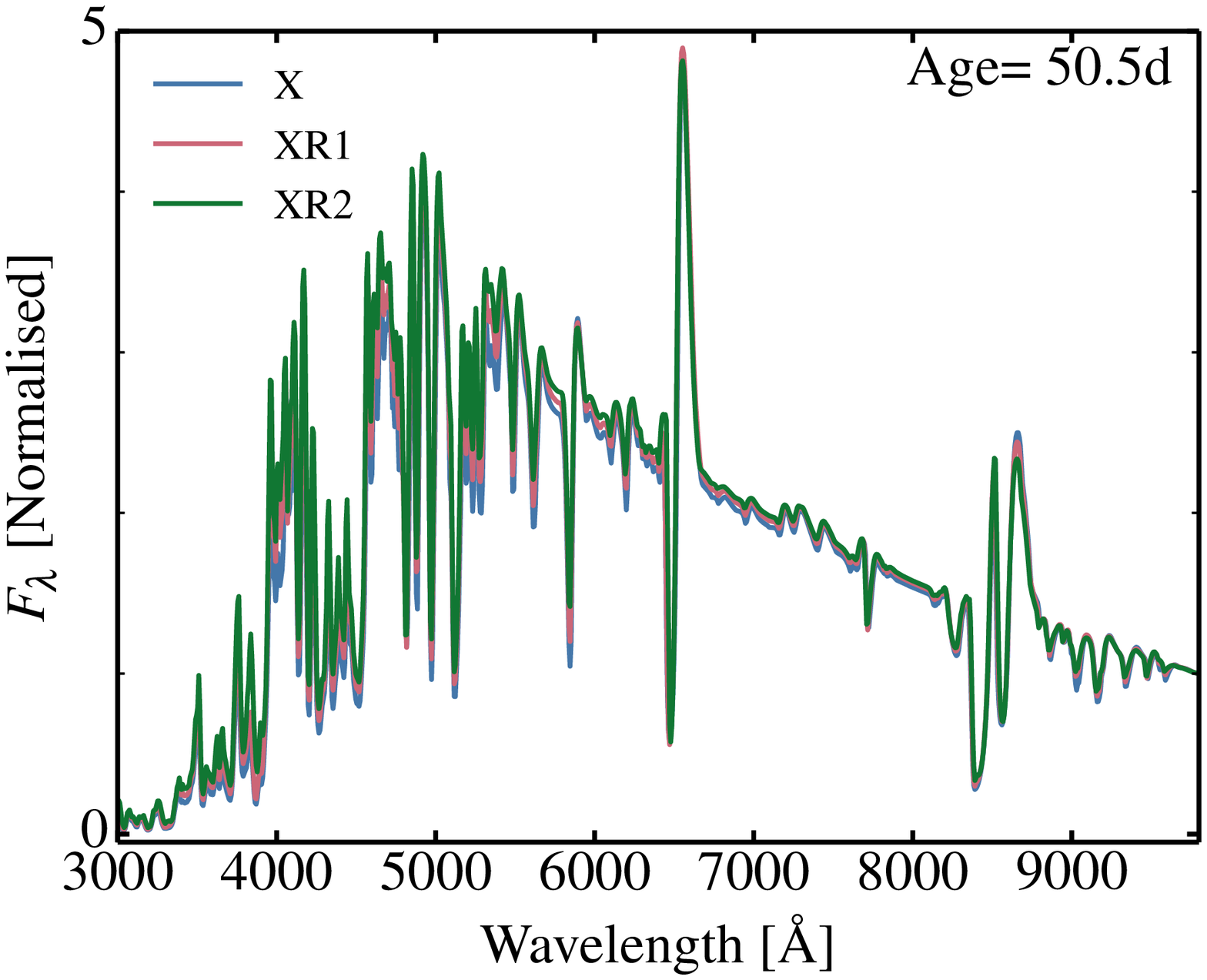}
    \caption{
    Comparison between synthetic spectra (normalised at 9800\,\AA) of models X, XR1, and XR2 at the onset of
    the recombination phase (13.3\,d; top) and half-way through the plateau phase (50.5\,d; bottom).
    A change in progenitor radius visibly impacts the spectral appearance early on, but the
    impact is subtle at late times.
    \label{fig_spec_rad}}
\end{figure}

\section{Sensitivity to progenitor and explosion properties}
\label{sect_dep}

In this section, we study the impact of slight changes in the progenitor radius,
chemical mixing, and ejecta mass on the resulting
photometric and spectroscopic predictions.
In the process, we build physical error bars for the properties of
the progenitor and explosion associated with SN\,2008bk.


\subsection{Radius}
\label{sect_rad}

  To test the influence of a change in progenitor radius, we use the same approach as in \citet{dessart_etal_13}.
In practice, we vary the efficiency of convection within the MLT formalism in \mesa, by tuning the
parameter $\alpha_{\rm MLT}$ from 3 (model X), to 2.5 (model XR1), and 2.0 (model XR2).
As $\alpha_{\rm MLT}$ decreases, the RSG star model at death increases in radius, from 502 (X), 
to 581 (XR1), and 661\,\rsun\ (XR2).
The main effect of a change in the RSG progenitor radius is to mitigate the cooling of the ejecta 
as it expands.
The larger the progenitor star, the weaker the cooling of the ejecta.
Here, to isolate the influence of a change in $R_\star$, models X, XR1, and XR2 are exploded
to produce ejecta with about the same kinetic energy at infinity
($\approx$\,2.6\,$\times$10$^{50}$\,erg) and about the same \iso{56}Ni mass (0.008--0.009\,\msun) ---
see Table~\ref{tab_models} for details.

Figure~\ref{fig_lc_rad} shows the bolometric luminosity, the absolute $V$-band and the $(U-V)$ light curve
for models X, XR1, and XR2.
A larger progenitor radius yields a larger bolometric luminosity, a greater $V$-band brightness,
and a bluer colour throughout the photospheric phase. It does not affect the length of the plateau phase.
For the bolometric luminosity, the offset is roughly constant during the photospheric phase and
corresponds to an increase of 15\% (for an increase of 16\% in $R_\star$) between models X and XR1,
and an increase of 15\% (for an increase of 14\% in $R_\star$) between models XR1 and XR2.
The offset of about 0.2\,mag in $V$-band magnitude between X and XR1, and between XR1 and XR2,
is also roughly constant past 30\,d after explosion --- the offset is reduced at earlier times
in part because of the different colours of the model early on.
The colour is systematically bluer for the more extended progenitor star, and the more so at earlier times.
The offset in $(U-V)$ is about 0.2\,mag between each model for most of the photospheric phase, but it
decreases near the end. Consequently, changes in progenitor radii may introduce a scatter in the intrinsic colour of SNe II-P.

Figure~\ref{fig_spec_rad} compares the synthetic spectra for models X, XR1, and XR2 at 13.3\,d and 50.5\,d after explosion.
We normalise the models at 9800\,\AA\ to better reveal the relative variation in flux across the UV and optical.
The colour shift discussed above is visible in the spectra at early times.
The bluer spectrum of model XR2 at a given post-explosion epoch is suggestive of a higher
photospheric temperature and ionisation. For example, the Ca\two\ near-IR triplet is hardly visible in model XR2
at 13.3\,d, but readily visible as a weak absorption feature in model X  
in which the transition to the recombination phase is more advanced.
As apparent, a change of only 15\% in the progenitor radius causes a significant difference in colours
at early times (this offset may be missed without the correct explosion time since a larger progenitor
radius merely shifts the colour  to later in time).
The spectral offset between models X, XR1, and XR2 is however smaller than obtained
between the models MLT1 and MLT3 of \citet{dessart_etal_13},
which resulted from the explosion of RSG stars with a much larger difference in surface radius (1100 and
500\,\rsun, respectively). Modulations in intrinsic colour may stem from diversity in progenitor radii,
although in observations, such variations can also originate from reddening.

  Model X is brighter than SN\,2008bk by 0.5\,mag during the photospheric phase. Given the results for
  models XR1 and XR2, a reduction
  in radius by 15\% would make it fainter by 0.2\,mag --- this could arise from the adoption of a lower metallicity,
  like for model Y.
  But a greater reduction would be needed to cancel the present offset and this would likely
  impact the colour. However, the colour of model X matches that of SN\,2008bk, and confirms the need
  for a rather compact RSG progenitor star, as advocated by \citet{dessart_etal_13} for SNe II-P in general
  (see also \citealt{gaitan_sn2_15}).
  To reconcile model X with the observations (brightness and expansion rate inferred from line profile widths)
  would likely require a lower ejecta kinetic energy (or an increase in ejecta mass). A 10-20\% increase
  in the distance to SN\,2008bk would reduce the offset, just like an increase in reddening from 0.02 to 0.1\,mag.

\begin{figure}
\includegraphics[width=0.47\textwidth]{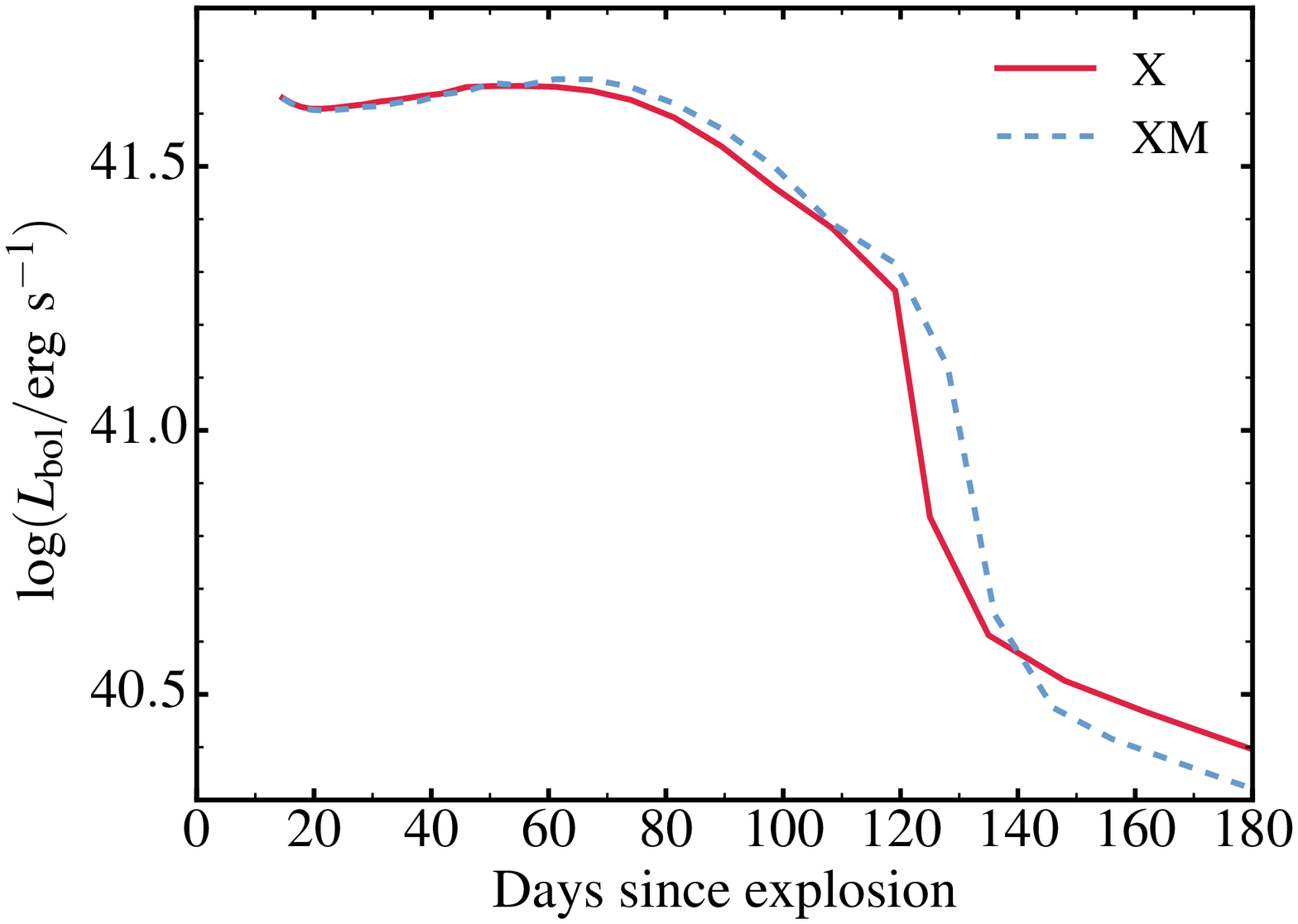}
\includegraphics[width=0.47\textwidth]{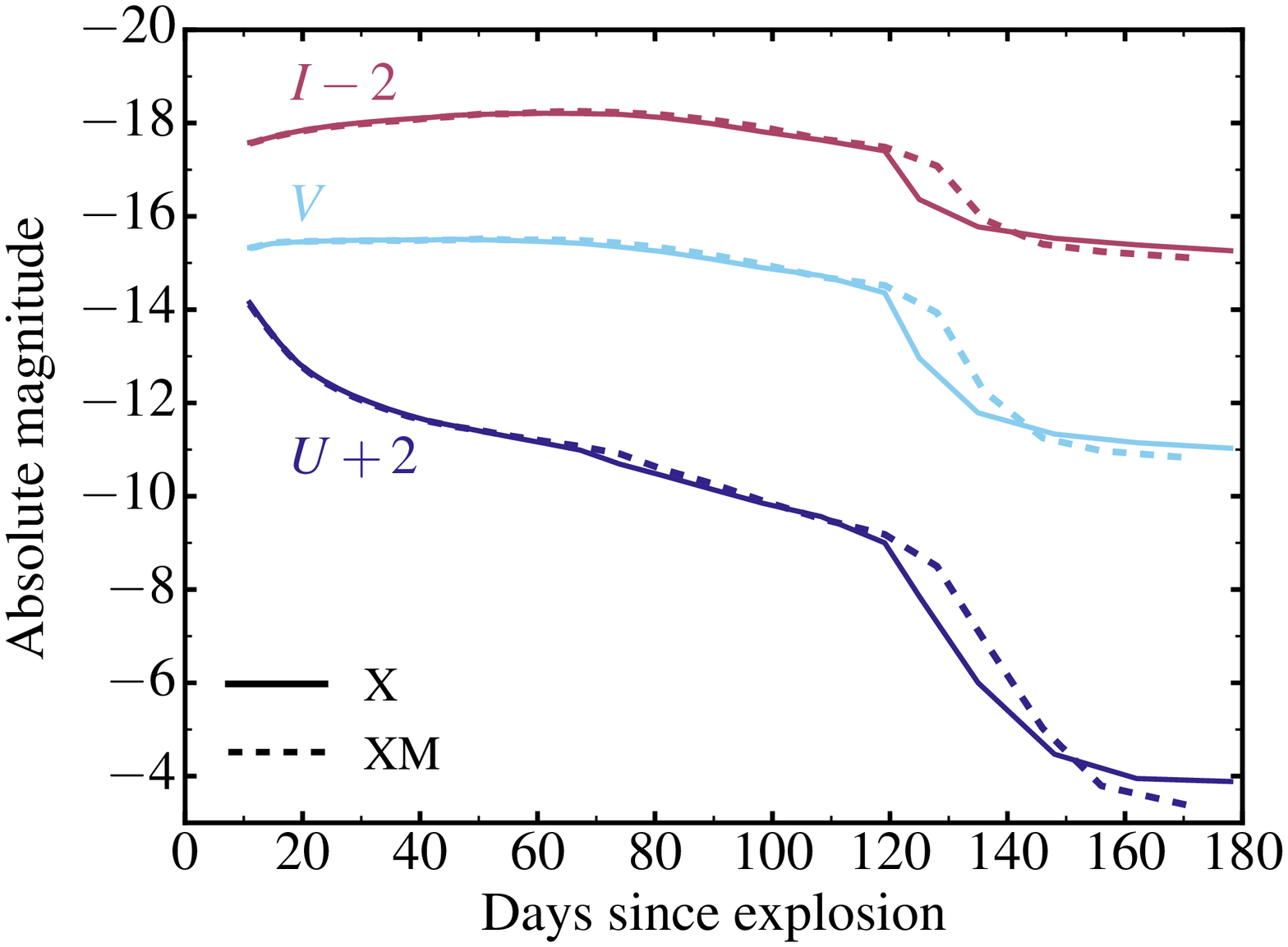}
\includegraphics[width=0.47\textwidth]{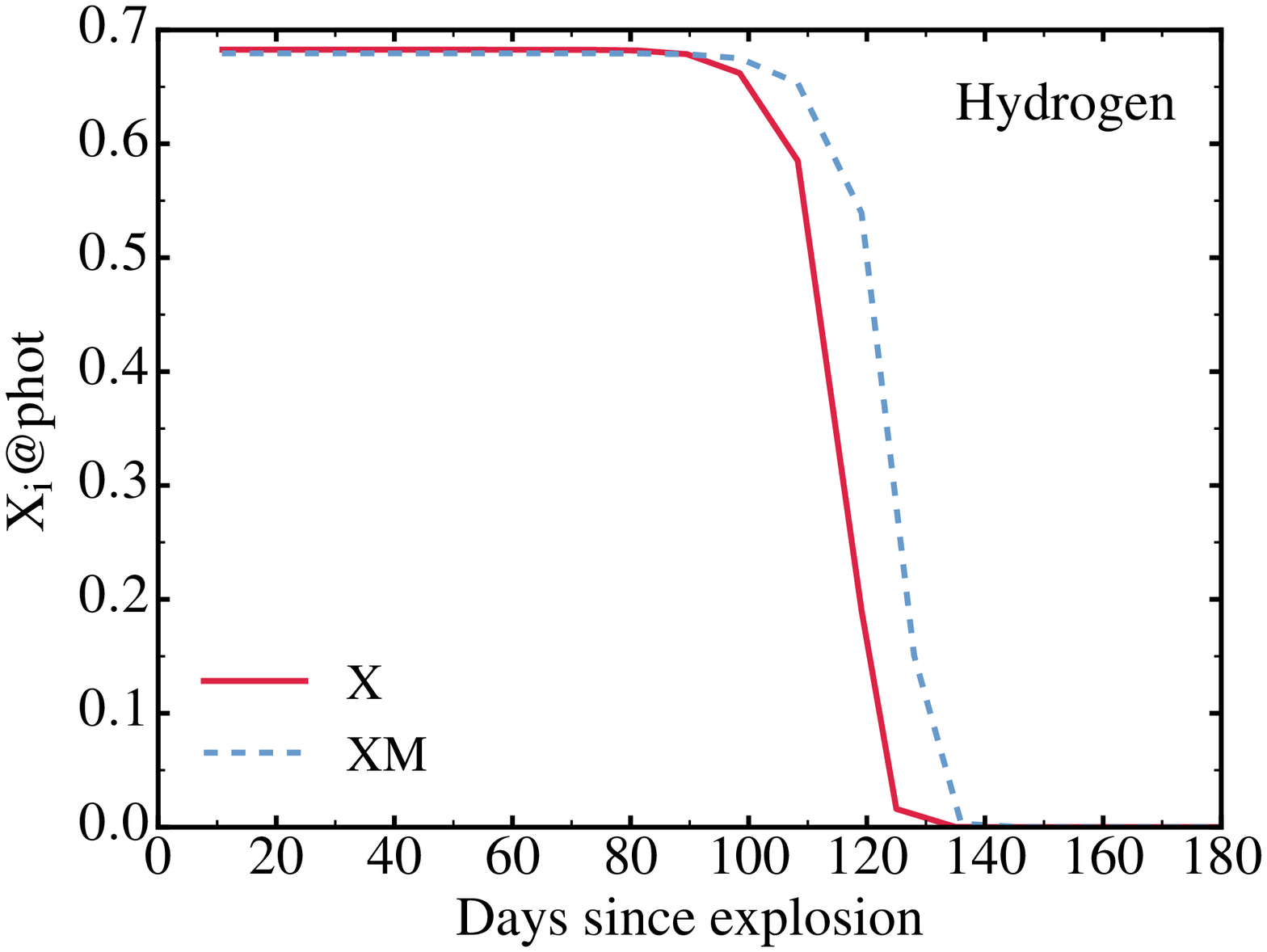}
\caption{
Comparison between models X and XM showing the bolometric light curve
(top), the $UVI$ light curves (middle), and the H mass fraction at the photosphere
during the first 180\,d after the explosion.
   \label{fig_mass}}
\end{figure}


\subsection{Mass}
\label{sect_mass}

   In this section, we discuss the impact of a different ejecta mass on the resulting
SN radiation. We compare our best-match model X with model XM, which starts
on the main sequence with the same initial conditions (mass, metallicity, etc.)
but evolved with a smaller mass loss rate (see Section~\ref{sect_model}).
This influences primarily the RSG phase and the impact is limited to the H-rich envelope mass, which is 1\,\msun\
greater in model XM --- both models have the same He-core mass.
The corresponding SN ejecta reflect this difference (see Table~\ref{tab_models}).
In practice, the explosion models X and XM differ slightly in explosion energy
so that the $E_{\rm kin}/M_{\rm e}$ is about the same for both.

Figure~\ref{fig_mass} shows the resulting bolometric light curve, the $UVI$ light curves,
and the variation of the H mass fraction at the photosphere. The light curves are essentially
identical for models X and XM, except for the longer duration of the high-brightness
phase in model XM. Here, an offset in ejecta mass of 1\,\msun\ (which is tied to a difference
in the progenitor H-envelope mass) extends the photospheric phase
by $\sim$\,10\,d.\footnote{We do not show a spectral comparison because both models
are essentially identical at all epochs apart from the timing of the transition to the nebular phase.}
The bottom panel of Fig.~\ref{fig_mass} emphasises how the high-brightness phase coincides
with the epochs during which the photosphere is located within the layers that were formerly
part of the low-density H-rich progenitor envelope.

As the photosphere moves into the layers that were formerly in the highly-bound high-density
He core, the luminosity plummets. In our model X, when the optical depth at the base of the
H-rich layers drops to 1, the total electron-scattering optical depth across the deeper
layers (from the former He core) is only 10 and it drops
to 1 within a week. This whole transition occurs during the fall off from the plateau.
The light curve transition can be modulated by mixing of H and He. However, the properties
of the highly bound slow-moving He-core material are hard to constrain from the light curve since
the corresponding ejecta regions have a low/moderate optical depth at the end of the plateau phase.

\begin{figure*}
\includegraphics[width=0.48\textwidth]{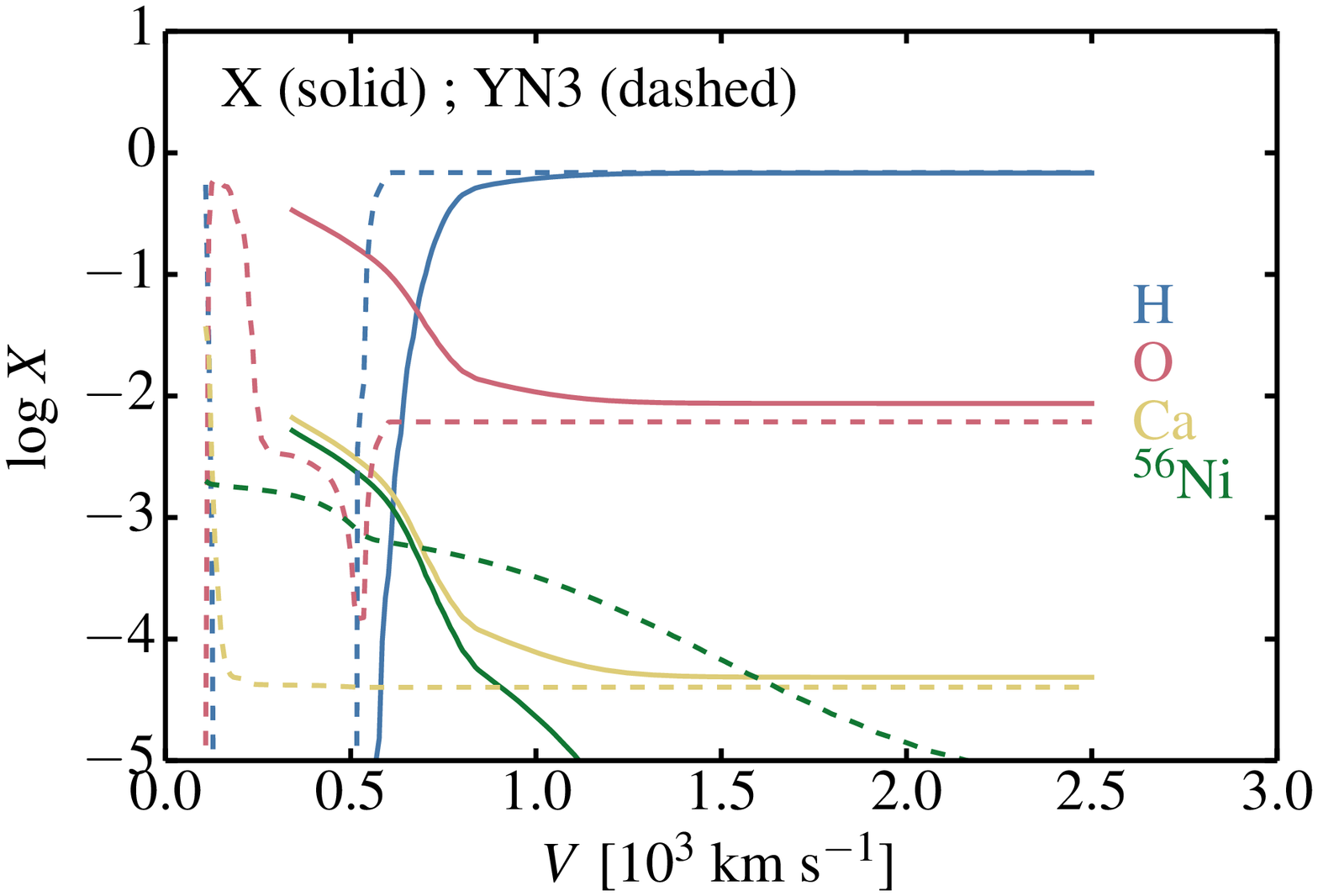}
\includegraphics[width=0.48\textwidth]{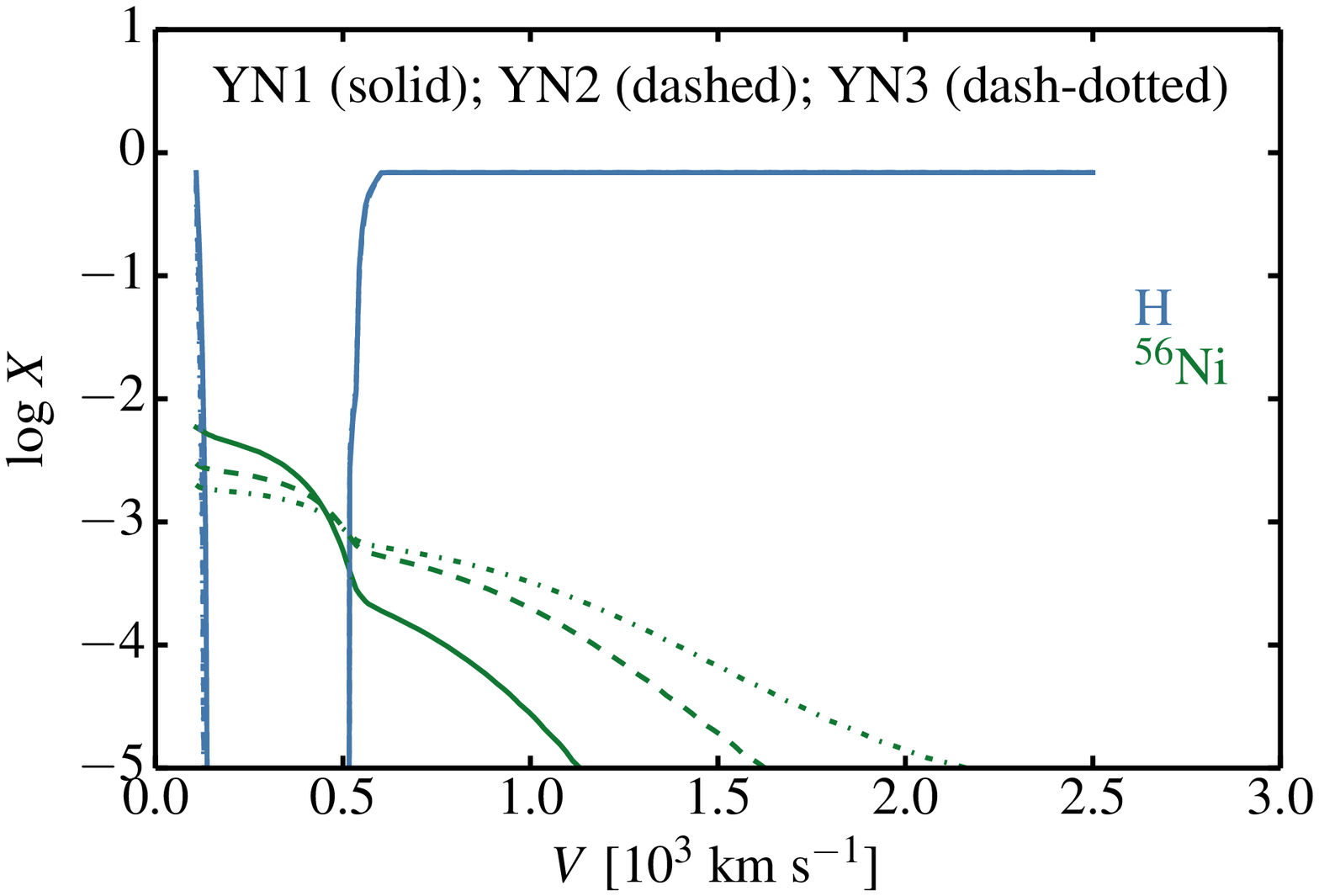}
\includegraphics[width=0.48\textwidth]{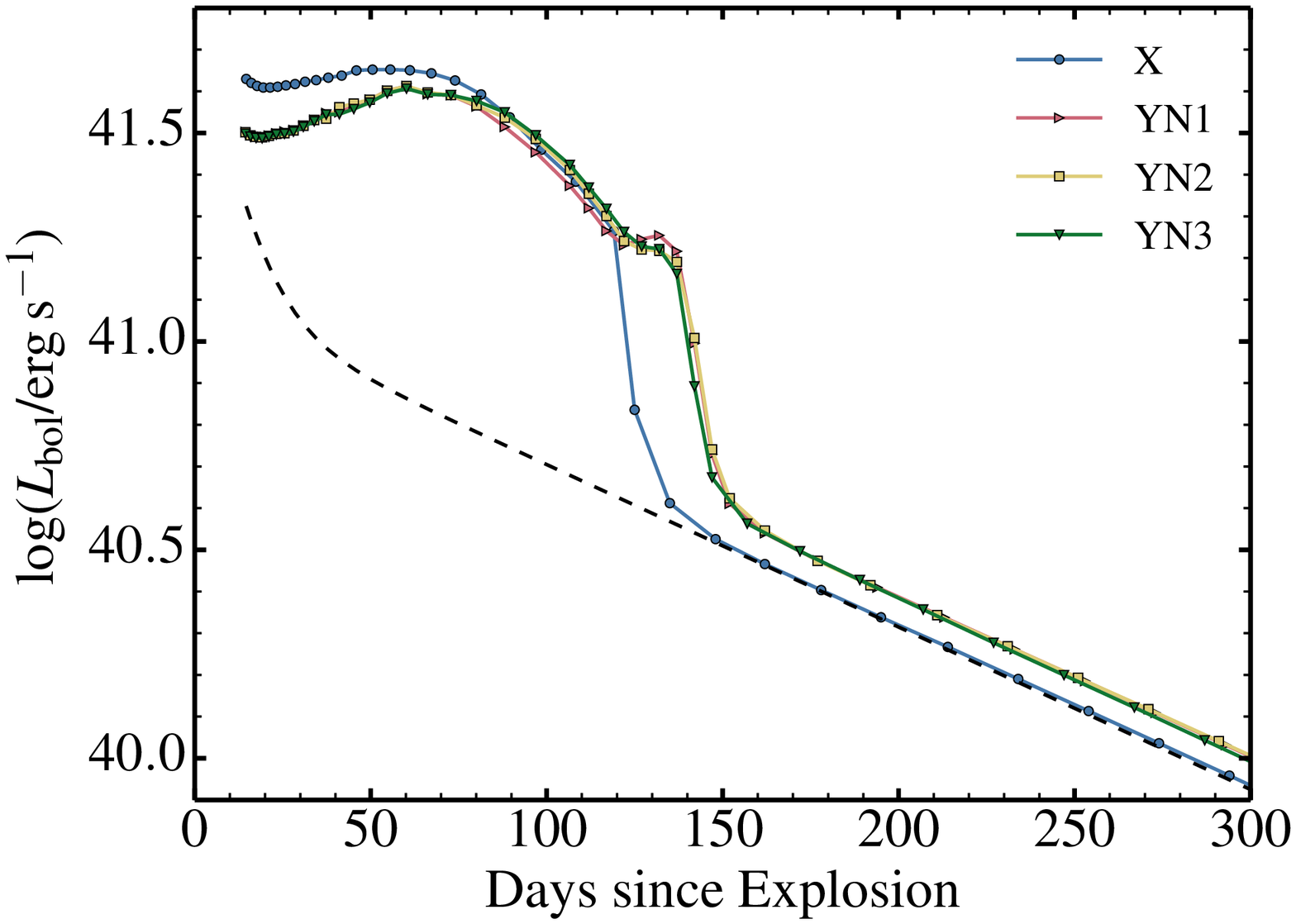}
\includegraphics[width=0.48\textwidth]{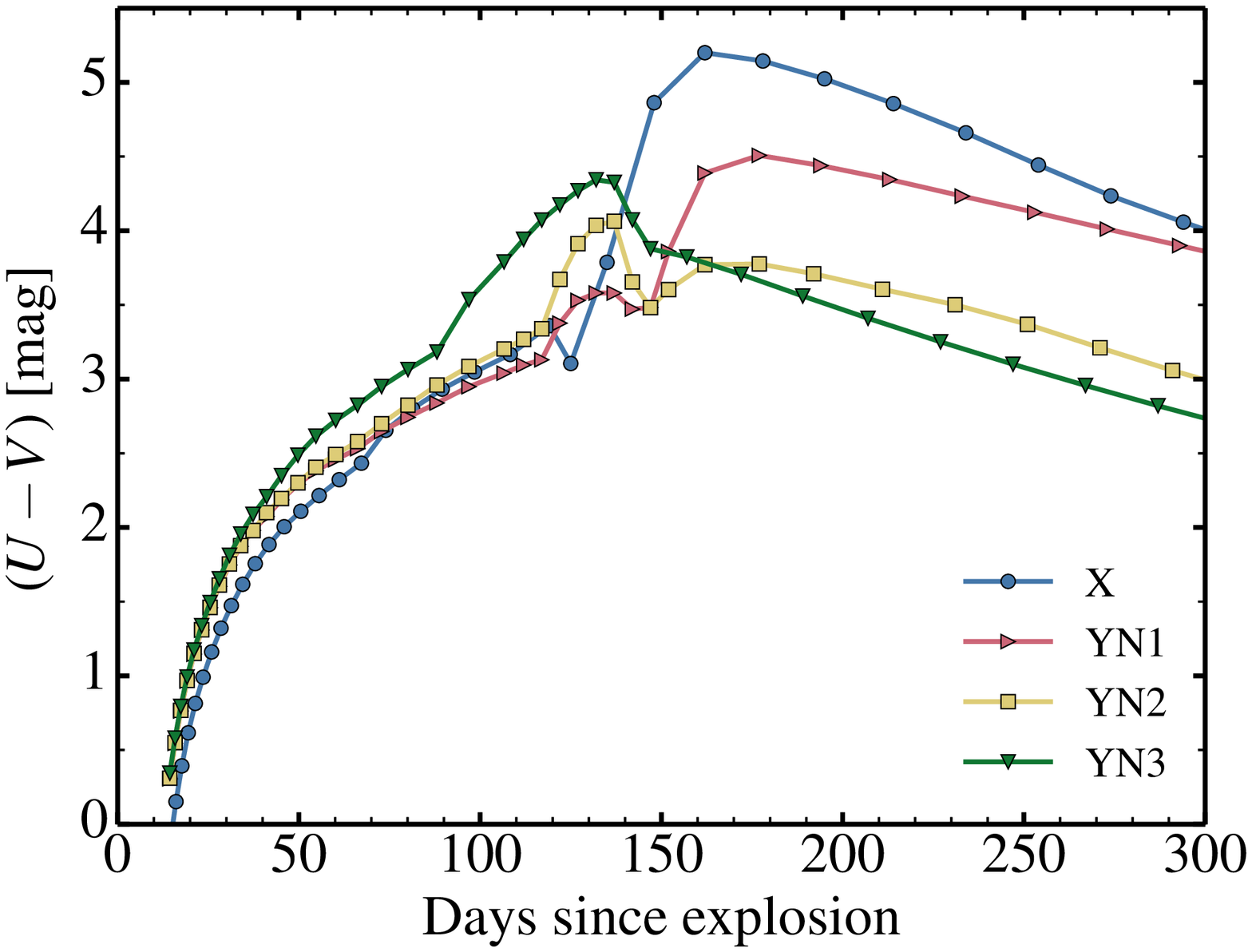}
\includegraphics[width=0.48\textwidth]{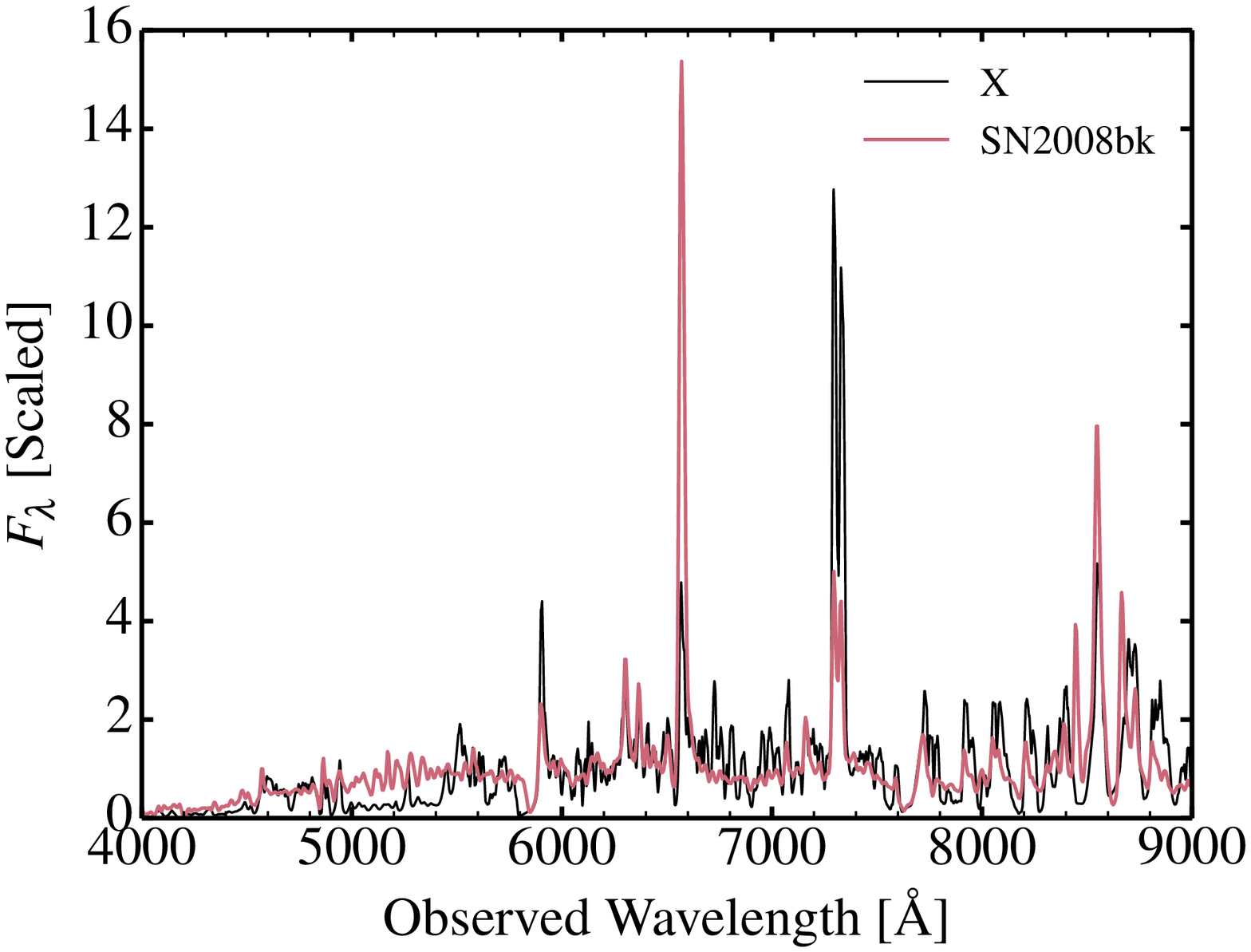}
\includegraphics[width=0.48\textwidth]{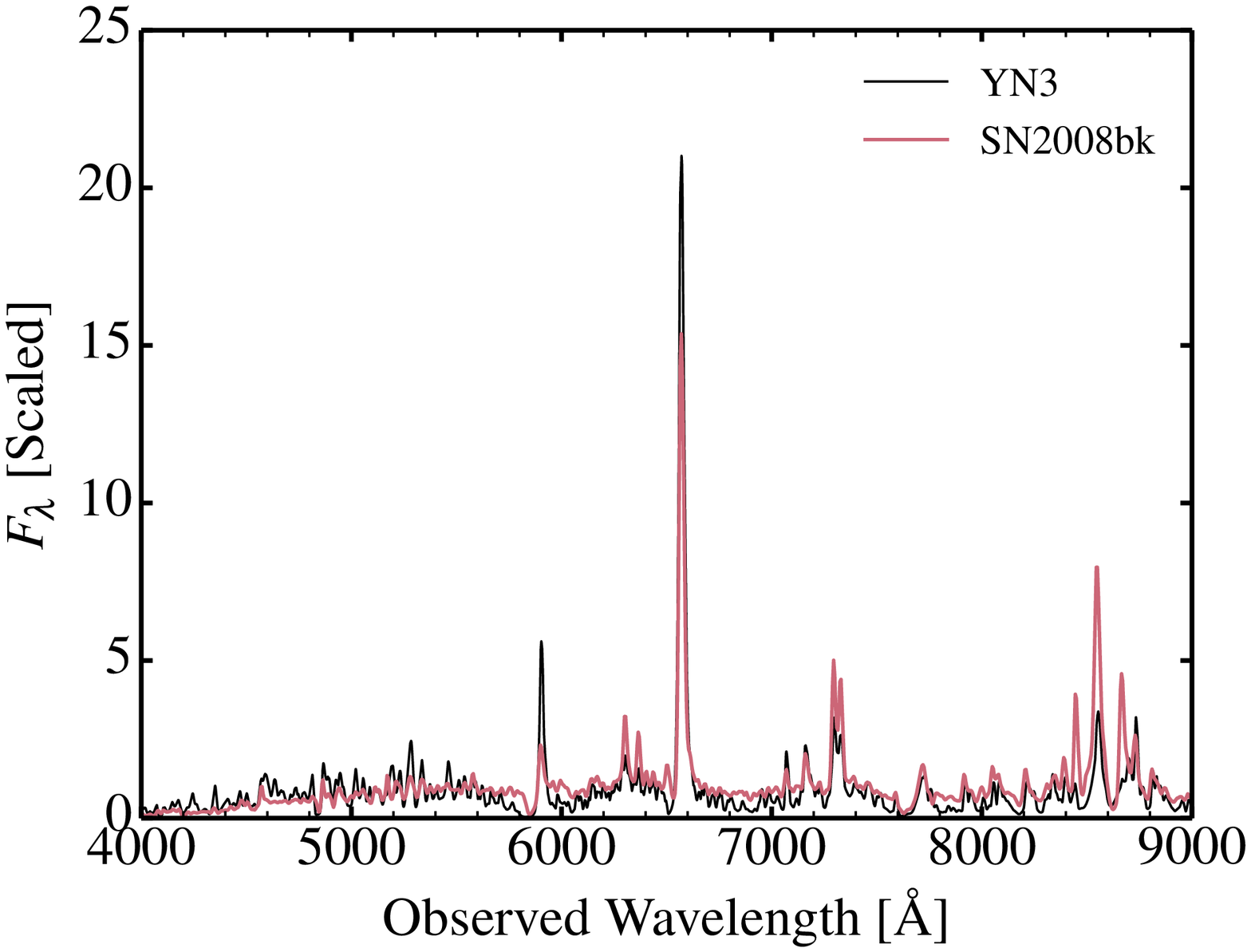}
\caption{
Top row: Composition profile for models X and YN3 (left),
and for models YN1, YN2, and YN3 (right) versus velocity (only the inner 2500\,\kms\
are shown to emphasize the properties of the inner ejecta regions where the effect of mixing is strong).
Middle row: Comparison of the bolometric luminosity (the dashed line gives the 
instantaneous decay power from an initial \iso{56}Ni mass of 0.00857\,\msun) and $(U-V)$ colour for models
X, YN1, YN2, and YN3. The longer photospheric phase for models YN1, YN2, and YN3 is
consistent with their progenitor having a $\sim$\,1\,\msun\ larger H-rich envelope mass.
Bottom row: Comparison of model X (left) and model YN3 (right) with the observations
of SN\,2008bk at 282\,d after explosion.
Model X underestimates the strength of H$\alpha$ but overestimates the strength of  the Ca\two\ doublet
at 7300\,\AA. In contrast, the strongly mixed model YN3 matches quite well Ca\two\,7300\,\AA\
and overestimates somewhat the strength of H$\alpha$.
\label{fig_mix}}
\end{figure*}

\begin{figure*}
\includegraphics[width=0.48\textwidth]{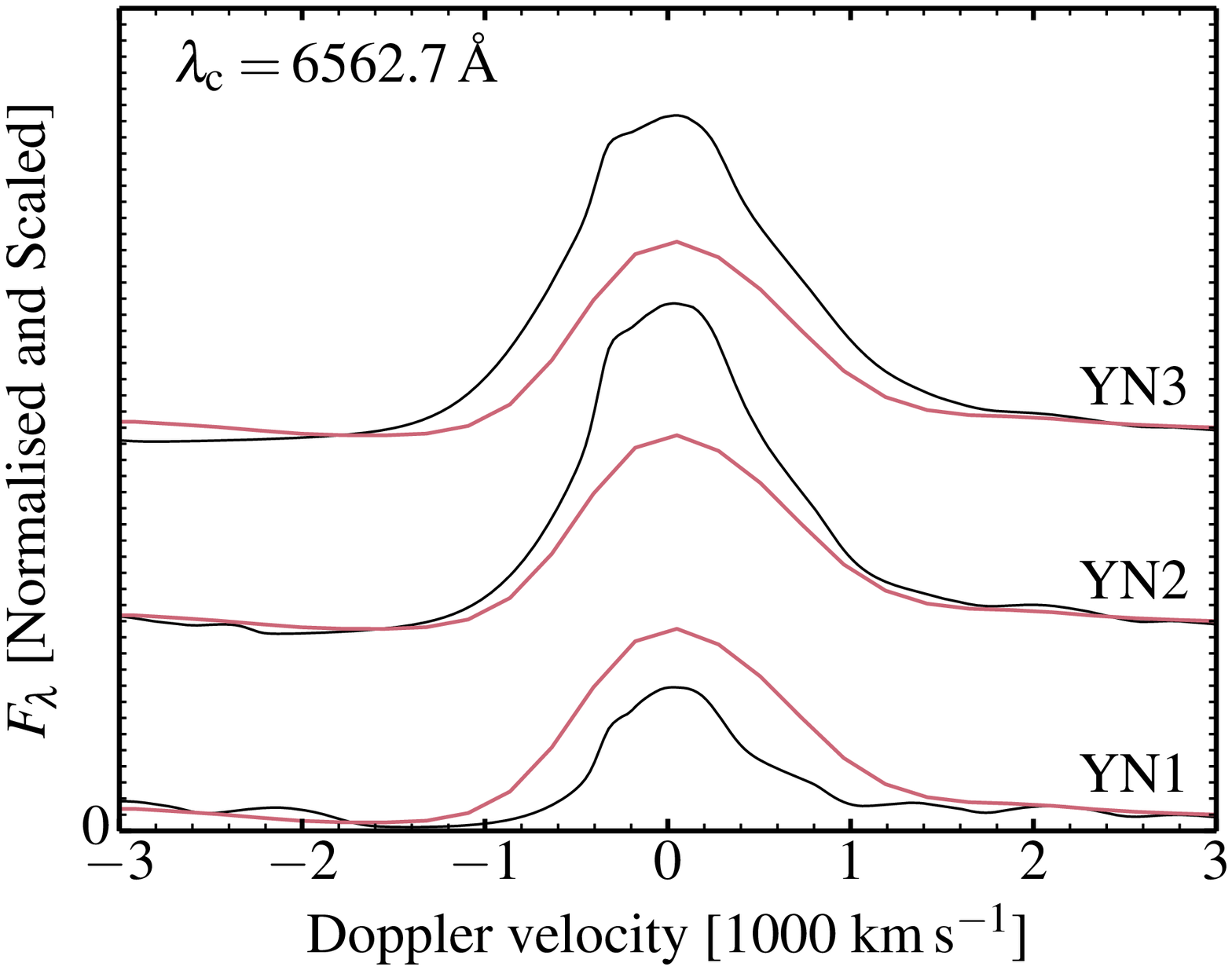}
\includegraphics[width=0.48\textwidth]{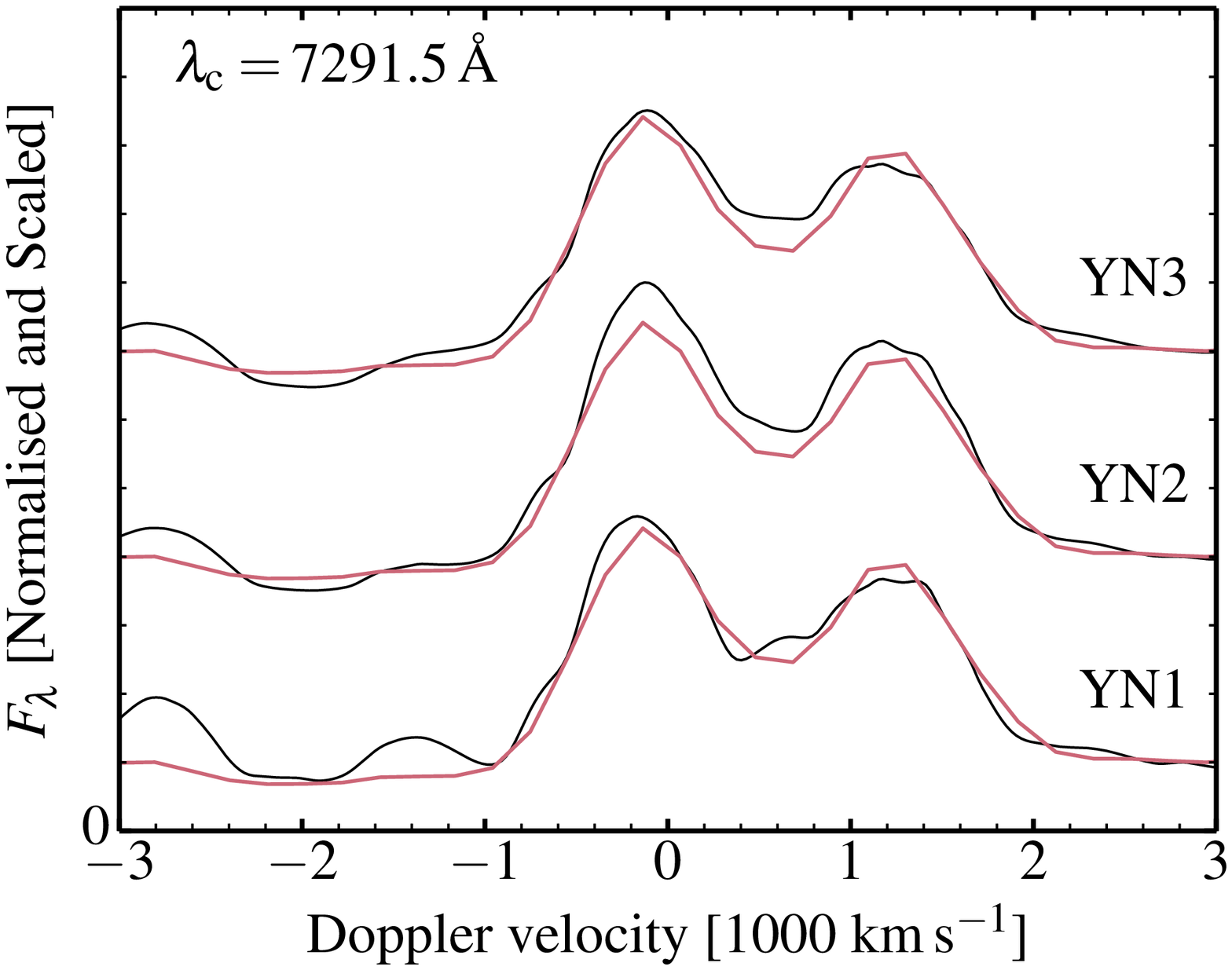}
\caption{
Comparison between the observations of SN\,2008bk at 282\,d after explosion (red)
with the models YN1, YN2, and YN3 (black) at the same epoch and for the spectral regions around
H$\alpha$ (left) and
the Ca\two\ doublet (the x-axis is centred on the rest wavelength of the blue component, at 7291.5\,\AA).
In all cases, we normalise all spectra at the red-most wavelength and shift vertically for better visibility.
\label{fig_YN_line}
}
\end{figure*}


\subsection{Mixing}
\label{sect_mixing}

We now discuss mixing in order to test how our implementation affects our results.
We use two approaches that are crude representations of the process
of mixing as simulated in core-collapse SNe \citep{mix91,wongwathanarat_15_3d}.
In the first approach, we step through each ejecta mass shell $m_i$ and mix all mass shells within
the range [$m_i$, $m_i+\delta m$] with $\delta m=$\,0.4\,\msun\ --- this is the usual way we proceed
\citep{dessart_etal_12}. The advantage is that it is straightforward to implement. This mixing is carried
out during the \v1d\ simulation, at 10,000\,s after the explosion was triggered (and thus about one day before
shock breakout). This mixing is both macroscopic (material is shuffled in mass/velocity space)
and microscopic (each 1-D/spherical mass shell on the \cmfgen\ grid is homogeneous). This is the mixing done
in models X, XR1, XR2, and XM.

In the second approach only \iso{56}Ni is mixed; all other species
are essentially left as in the original \mesa\ model except for hydrogen.
In practice, after mixing  \iso{56}Ni, regions where the sum of mass fractions is below unity,
hydrogen is added, while in regions where the sum of mass fractions is above unity (typically by 1\%),
all species' mass fractions are scaled so that the total is unity. This causes the spike in the H mass fraction
in the innermost shells of models YN1, YN2, and YN3 because the \iso{56}Ni mass fraction was close
to unity in those layers prior to mixing (Fig.~\ref{fig_mix}). This mixing approach is used
in models YN1, YN2, and YN3, which differ in the values of $\delta m$ used. In this order,
we have used $\delta m=$\,0.5, 1.0, and 1.5\,\msun.
We show the composition for a selection of species in the top row of Fig.~\ref{fig_mix}.

This exercise is instructive because our various models all derive from a 12\,\msun\ main sequence
star, with similar H-rich envelope and He-core properties. Yet, these different levels of mixing
produce drastic variations at nebular times, as we discuss now.

The middle-left panel of Fig.~\ref{fig_mix} shows the bolometric luminosity for models X and YN1/YN2/YN3.
During the photospheric phase, the early-time luminosity is higher in model X because of the large
progenitor radius (502 compared to 405\,\rsun) and the higher ratio $E_{\rm kin}/M_{\rm e}$.
The lower ejecta mass of model X (8.29 compared
to 9.45\,\msun) produces a shorter plateau length by 20\,d (all four models have the same ejecta kinetic
energy of $2.5\times 10^{50}$\,erg).
Interestingly, the three models YN1/YN2/YN3 show a similar bump at the end of the plateau phase, despite
their different levels of  \iso{56}Ni mixing.
Model X, which has a similar level of  \iso{56}Ni mixing as model YN1, does not show a bump.
This bump is controlled by the mixing of other species, and in particular how
H and He are mixed in velocity/mass space at the He-core edge (see, e.g., \citealt{utrobin_99em_07}).
In model X, all species are
mixed and this tends to soften the changes in composition in the inner ejecta, which the photosphere
probes at the end of the plateau phase. H is important here because it is the main electron donor
so that different levels of H mixing can modulate the evolution of the optical depth at the end of the
plateau phase (and modulate the recession of the photosphere and the variation in luminosity).

The middle-right panel of Fig.~\ref{fig_mix} shows the evolution of the $(U-V)$ colour for the four models.
The more extended model X is bluer during the photospheric phase, but redder at nebular epochs. 
Model YN3 with the strongest mixing is the redder of all four models
during the photospheric phase -- the extra heating does not compensate for the increased opacity from
metals, but it is the bluer of all models at nebular times. 
The drastic change is seen when the models transition to the nebular phase.
Model X becomes exceedingly red, while models YN1, YN2, and YN3 remain systematically bluer, the more
so the larger the mixing of \iso{56}Ni.

The broad-band fluxes of Type II SNe do not depend on the temperature in the same way at nebular times as
during the photospheric phase, when the escaping radiation has roughly the properties of a blackbody.
At nebular times, the SN radiates
the energy deposited by radioactive decay through strong emission lines,
many of which being forbidden. Here, temperature and ionisation of the gas
control which lines cool the ejecta. Then, depending on the location of these transitions in wavelength space,
one can produce different colours, even if the bolometric luminosity (which is set by the amount of decay energy
absorbed by the gas) is the same between models.

In the bottom row of Fig.~\ref{fig_mix}, we show a comparison of model X (left) and model YN3 (right)
with the observations of SN\,2008bk at 282\,d after explosion (the models and observations have been
normalised at 6310\,\AA\ to better reveal the relative offsets in flux).
Model X follows roughly the observed spectral energy distribution, but it underestimates the flux in the 5000\,\AA\ region
and overestimates it in the red part of the optical.
Model X has cooler and more recombined H-rich shells (e.g., with dominance of Fe\one) than model YN3 (with dominance of Fe\two).
More importantly, it strongly overestimates the Ca\two\,7300\,\AA\ doublet
and underestimates the strength of H$\alpha$.
In contrast, model YN3, which has the same amount of \iso{56}Ni and a similar core composition as model X,
has a completely different nebular spectrum.
Model YN3 matches well the Ca\two\,7300\,\AA\ doublet and the Ca\two\,8500\,\AA\ triplet
and overestimates the strength of H$\alpha$ by about a factor of two (model YN1, characterised
by a weaker \iso{56}Ni mixing, slightly underestimates the strength of H$\alpha$).
The half-width-at-half-maximum is $\sim$\,500\,\kms\ for H$\alpha$  (this velocity corresponds to the innermost H-rich layers
in the ejecta) and is about 350\,\kms\ for the Ca\two\,7300\,\AA\ doublet (this velocity corresponds to the He-rich ejecta regions).
In both cases, the line forms over a range of velocities. For H$\alpha$, model YN3 strongly overestimates the extent
of the absorption trough --- model YN1 with a weaker \iso{56}Ni mixing has a narrower trough that matches closer
the observations (Fig.~\ref{fig_YN_line}). The different levels of mixing affect the H$\alpha$ widths,
but not Ca\two\,7300\,\AA\ --- the \iso{56}Ni mixing affects more the ionisation of H\one\ than that of Ca\two\
in the outer ejecta.

This short exploration on mixing highlights the complications it introduces even for ejecta that have the same overall
composition, mass, and energetics. How the different elements are distributed at both the microscopic and macroscopic level is
a fundamental aspect of the problem \citep{mix91,jerkstrand_etal_12,wongwathanarat_15_3d},
and it depends both on the composition prior to explosion as well on the impact of the explosion on this distribution.
We will come back to the issue of mixing for nebular phase spectra in a future study.

\section{Conclusions}
\label{sect_conclusion}

    We have presented the results of numerical simulations that aim at understanding the properties
of the ejecta/radiation associated with SN\,2008bk, as well as the progenitor star at its origin.
We have focused on a single mass of 12\,\msun\ for the progenitor, and have evolved this model
with \mesa\ at solar metallicity but with different parameterisations for mass loss and convection in
order to produce RSG stars at death that cover a range of final masses and surface radii.
These  models were then exploded with the radiation hydrodynamics code \v1d\ to produce
ejecta with various explosion energies and \iso{56}Ni mass. Finally, starting at a post-explosion time
of 10\,d, we have evolved these ejecta with the nLTE radiative transfer code \cmfgen, building
multi-band light curves and multi-epoch spectra that can be directly compared to observations.

  Our model X, which closely matches SN\,2008bk, corresponds to a star with an initial mass of 12\,\msun,
a final mass of 9.88\,\msun, an H-rich envelope mass of 6.7\,\msun\, a final surface radius of 502\,\rsun.
The associated ejecta has a mass of 8.29\,\msun, a kinetic energy of 2.5$\times$\,10$^{50}$\,erg and
0.0086\,\msun\ of \iso{56}Ni.
Provided we introduce a 0.5\,mag offset, model X follows closely the multi-band optical and near-IR
light curves of SN\,2008bk, including the colour evolution (e.g., the sharp drop in the $U$ band and the
near-constant evolution of the $V-$band magnitude)  and the plateau duration.
Model X reproduces well the spectral evolution of SN\,2008bk (provided we renormalise the spectra
to cancel the 0.5\,mag offset in brightness), the progressive reddening of the spectra as the photospheric
temperature drops and line blanketing strengthens, the reduction in line widths as the photosphere recedes
to deeper/slower ejecta layers. The early colour evolution and reddening of the spectra is best matched
with an explosion date of MJD\,54546.0.
We find that model X yields lines that are 10-20\% too broad --- it
typically overestimates the expansion rate by 10--20\%.
Our model X is therefore somewhat too energetic for SN\,2008bk. A lower explosion energy could reduce
the offset in line widths and brightness. The offset in luminosity could be reduced
{by invoking a larger reddening (perhaps up to $E(B-V)=$\,0.1\,mag) to SN\,2008bk.
The uncertainty in the Cepheid-based distance to the host of SN\,2008bk is probably small.}
Given all the uncertainties involved, these offsets
are reasonably small to suggest that model X is a sensible representation of SN\,2008bk.

Our model of SN\,2008bk is a low-energy counterpart of the models for ``standard" SNe II-P like 1999em.
The mechanisms that control their evolution are the same for both --- in our approach we merely reduce
the energy injection to produce a model for SN\,2008bk rather than one for SN\,1999em.
However, the lower expansion rate in SN\,2008bk allows a much better inspection of the line profile fits.
Interestingly, in low-luminosity SNe II, the H$\alpha$ line systematically develops a complex structure
at the end of the plateau phase. This structure is most likely not a signature of asymmetry, but instead
caused by overlap with the strong Ba\two\,6496.9\,\AA\ line. This conclusion is reinforced by the good
match obtained to the isolated line of Ba\two\,6141.7\,\AA\ in SN\,2008bk.

   We have performed additional models to gauge the sensitivity of our results to changes in progenitor/explosion
parameters. A 15\% increase in the progenitor surface radius leads to a $\sim$\,15\% increase in plateau
luminosity and a 0.2\,mag brightening of the $V$-band magnitude, but does not affect the length of the photospheric
phase. An increase in ejecta mass of 1\,\msun\ hardly affects the results for model X, except for the lengthening
of the photospheric phase by 10.0\,d.

    In our simplistic approach, chemical mixing has little impact on the SN radiation throughout
    most of the plateau phase.
Early on, increased mixing causes extra line blanketing and produces redder optical colours.
In our models where only \iso{56}Ni is mixed outwards (but no mixing is applied to H and He), the
light curve develops a 15-d long ledge before dropping from the plateau. Mixing of H with He  smoothes
that transition, most likely because it allows a much smoother evolution of the electron-scattering optical
depth, which controls the release of radiative energy stored in the ejecta (and the rate of recession
of the photosphere).
However, the most drastic impact of mixing in our models is seen at the nebular times. Depending on how
we perform the mixing, we can completely quench H$\alpha$, boost Ca\two\,7300\,\AA, or
mitigate the temperature and ionisation state of the gas to alter the SN colours.
Because of all these complications, we defer the modelling of the nebular-phase spectra of SN\,2008bk.

\section{Acknowledgements}
Sergey Lisakov is supported by the Erasmus Mundus Joint Doctorate Program by
Grants Number 2013-1471 from the agency EACEA of the European Commission.
LD acknowledges financial support from the Agence Nationale de la Recherche
grant ANR-2011-Blanc-SIMI-5-6-007-01.
DJH acknowledges support from NASA theory grant NNX14AB41G.

\appendix

\section{Line identifications for model X at early and late times in the photospheric phase}

\begin{figure*}
 \includegraphics[width=0.48\textwidth]{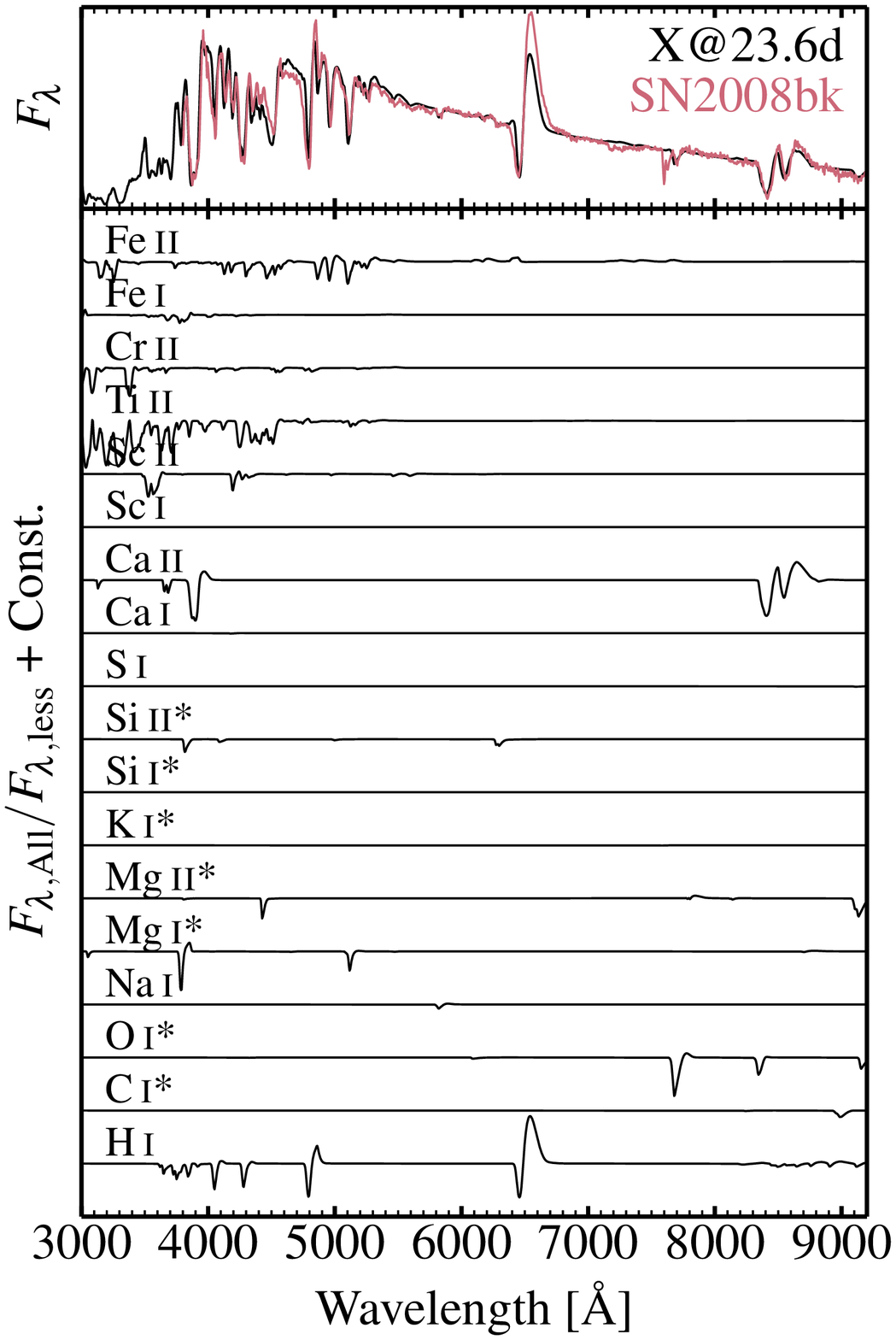}
  \includegraphics[width=0.48\textwidth]{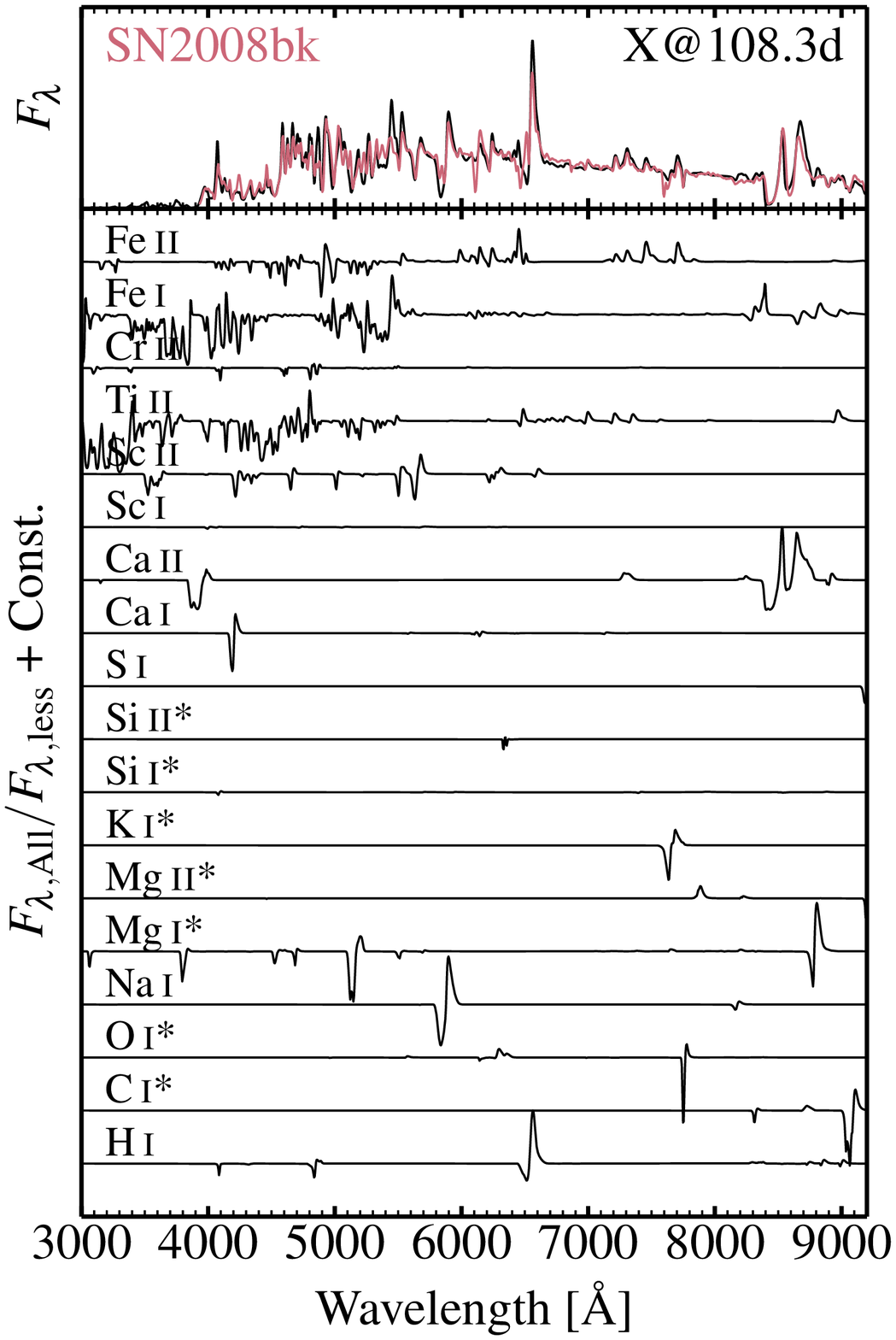}
  \vspace{-1cm}
\caption{Left: the top panel compares model X at 23.6\,d
with the observations (dereddened, deredshifted and normalised at 8000\,\AA)
of SN\,2008bk on 12th of April 2008 (which is about 23\,d after our adopted time of explosion
of MJD\,54546.0). The bottom panel stacks the quantity
$F_{\lambda,{\rm All}}/F_{\lambda,{\rm less}}$, where $F_{\lambda,{\rm All}}$ is the total
synthetic spectrum and $F_{\lambda,{\rm less}}$ is the spectrum computed with the bound-bound
transitions of one atom/ion omitted (see label). For starred species, we apply a scaling of 3 to reveal
the weak line features.
Right: Same as left, but now for model X at 108.3\,d after explosion and the observations of
SN\,2008bk on the first of July 2008 (which corresponds to 103.0\,d after our adopted time of explosion
of MJD\,54546.0; data from VLT/FORS spectropolarimetry program; Leonard et al. in prep).
Because the observed flux was not well calibrated, we distort the global shape of the observed spectrum to
match that of the model, which is fine here since the purpose is to inspect how line features compare
between model and observations.
\label{fig_X_ladder}}
  \end{figure*}

\label{lastpage}

\end{document}